\documentclass[conference]{IEEEtran}

\def\BibTeX{{\rm B\kern-.05em{\sc i\kern-.025em b}\kern-.08em
    T\kern-.1667em\lower.7ex\hbox{E}\kern-.125emX}}

\usepackage{cite}
\usepackage{amsmath,amssymb,amsfonts}
\usepackage{algorithmic}
\usepackage{graphicx}
\usepackage{textcomp}
\usepackage{xcolor}

\usepackage{todonotes}
\usepackage{listings}
\usepackage{ctable}
\usepackage{pifont} 
\usepackage{graphicx} 
\usepackage{seqsplit}
\usepackage{hyperref}

\usepackage[utf8]{inputenc}
\usepackage{tabularx}   
\usepackage{booktabs}   
\usepackage{float} 
\usepackage[most]{tcolorbox}
\usepackage{url}
\usepackage{hyperref}
\usepackage{multirow}
\usepackage{makecell}

\usepackage{soul}
\usepackage{annotate-equations}
\usepackage{ipdps26repro}

\definecolor{dkgreen}{RGB}{0,64,0}
\definecolor{ltgray}{RGB}{245,245,245}
\definecolor{mauve}{RGB}{139,0,139}

\definecolor{color_1}{HTML}{009E73} 
\definecolor{color_2}{HTML}{D55E00} 
\definecolor{color_3}{HTML}{0072B2} 
\definecolor{color_4}{HTML}{000000} 
\definecolor{color_5}{HTML}{E69F00} 
\definecolor{color_6}{HTML}{CC79A7} 

\colorlet{soul_1}{color_1!20}
\colorlet{soul_2}{color_2!20}
\colorlet{soul_3}{color_3!20}
\colorlet{soul_4}{color_4!20}
\colorlet{soul_5}{color_5!20}
\colorlet{soul_6}{color_6!20}

\newcommand{\tweakedsim}{\raise.17ex\hbox{$\scriptstyle\mathtt{\sim}$}}

\definecolor{belizehole}{HTML}{2980b9}
\newtcolorbox{RQcallout}[1][]{%
    colback=black!5,
    colframe=black!5,
    notitle,
    sharp corners,
    borderline west={2pt}{0pt}{belizehole!80!black},
    enhanced,
    breakable,
  }

\definecolor{myblue}{RGB}{63, 90, 126}
\definecolor{mygray}{RGB}{228, 244, 247}

\newcounter{obsnum}
\newcommand{\observe}[2]{%
  \refstepcounter{obsnum}\label{#1}%
  \begin{tcolorbox}[
    colback=mygray,
    colframe=myblue,
    title=Observation \ref{#1}%
  ]%
    #2%
  \end{tcolorbox}%
}

\makeatletter
\newcommand{\linebreakand}{%
  \end{@IEEEauthorhalign}
  \hfill\mbox{}\par
  \mbox{}\hfill\begin{@IEEEauthorhalign}
}
\makeatother

\begin{document}

\title{The Big Send-off: Scalable and Performant Collectives for Deep Learning}

\author{\IEEEauthorblockN{Siddharth Singh}
\IEEEauthorblockA{\textit{Department of Computer Science}\\
\textit{University of Maryland}\\
College Park, USA\\
ssingh37@umd.edu
}
\and
\IEEEauthorblockN{Keshav Pradeep}
\IEEEauthorblockA{\textit{Department of Computer Science}\\
\textit{University of Maryland}\\
College Park, USA\\
keshprad@umd.edu
}
\and
\IEEEauthorblockN{Mahua Singh}
\IEEEauthorblockA{\textit{Dept.~of Comp.~Sci.~and Engg.}\\
\textit{Indian Institute of Technology}\\
Guwahati, India\\
s.mahua@iitg.ac.in
}
\linebreakand
\IEEEauthorblockN{Cunyang Wei}
\IEEEauthorblockA{\textit{Department of Computer Science}\\
\textit{University of Maryland}\\
College Park, USA\\
cunyang@umd.edu
}
\and
\IEEEauthorblockN{Abhinav Bhatele}
\IEEEauthorblockA{\textit{Department of Computer Science}\\
\textit{University of Maryland}\\
College Park, USA\\
bhatele@cs.umd.edu
}
}

\maketitle

\begin{abstract}
Collective communication is becoming increasingly important in data center and
supercomputer workloads with an increase in distributed AI related jobs.
However, existing libraries that provide collective support such as NCCL, RCCL,
and Cray-MPICH exhibit several performance and scalability limitations on
modern GPU supercomputers.  To address these challenges, we introduce the
Performant Collective Communication Library (PCCL), specifically targeted for
distributed deep learning (DL) workloads.  PCCL provides highly optimized
implementations of key collectives used in distributed DL: all-gather,
reduce-scatter, and all-reduce. PCCL uses a hierarchical design with
learning-based adaptive selection of the best performing algorithms to scale
efficiently to thousands of GPUs. It achieves substantial performance speedups
over RCCL on 2048 GCDs of Frontier -- up to 168$\times$ for reduce-scatter,
33$\times$ for all-gather and 10$\times$ for all-reduce. More modest but still
significant gains up to 5.7$\times$ over NCCL are observed on Perlmutter.
These gains translate directly to performance improvement of
production DL workloads: up to 4.9$\times$ speedup over RCCL in DeepSpeed
ZeRO-3 training, and up to 2.4$\times$ speedup in DDP training.

\end{abstract}


\section{Motivation}
\label{sec:intro}
Communication overheads in parallel applications become an increasing fraction
of the overall execution time as these applications are scaled to more and more
nodes and GPUs. In the case of deep learning (DL)
applications~\cite{sc2020zero, fsdp, singh:ipdps2022, pytorchdist-vldb,
megatronlm, singh:ics2023, singh:ipdps2023, ranjan:sc2025}, communication is typically composed of collective operations such
as all-gather, all-reduce, and reduce-scatter. MPI and vendor-specific
libraries such as NCCL from NVIDIA and RCCL from AMD provide implementations of
such operations for use in DL frameworks. As opposed to traditional high
performance computing (HPC) applications, the message sizes in DL applications
are significantly larger, from tens to hundreds of MBs, and current
implementations of collective libraries struggle with maintaining high
performance in this regime.

In this work, we analyze the efficiency and scaling limitations of existing
communication libraries such as Cray-MPICH, NCCL and RCCL on multiple
platforms.  We examine three collectives that predominantly make up the
communication in DL workloads: all-gather, reduce-scatter, and all-reduce. We
discover that there are unique shortcomings, inefficiencies, and scalability
issues in each of the messaging libraries. Primarily, we observe that all
libraries struggle with efficiency and scalability in regimes of moderate message sizes (tens of MBs) and large
GPU counts.

\begin{figure}[h]
  \centering  
  \includegraphics[width=\columnwidth]{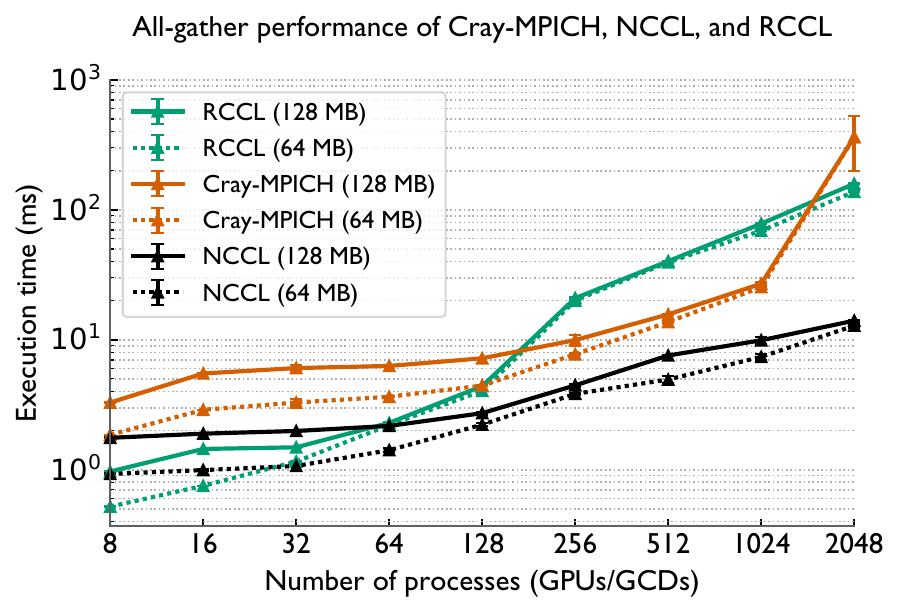}
  \caption{Performance comparison of all-gather using RCCL (Frontier), Cray-MPICH (Frontier), NCCL (Perlmutter)
  for 64 and 128 MB output buffer sizes.  The ideal scaling
behavior (flat horizontal line) is not achieved by either library, highlighting
their limited scalability at increasing GPU counts.}
  \label{fig:high-latencies}
\end{figure}

Figure~\ref{fig:high-latencies} presents the results of benchmarking the
all-gather operation on Frontier (AMD MI250X GPUs) and Perlmutter (NVIDIA A100
GPUs) with two output buffer (message) sizes: 64 and 128 MB using Cray-MPICH, NCCL, and RCCL.  We observed that Cray-MPICH is the slowest on lower
GPU/GCD counts but RCCL scales the most poorly beyond hundreds of GCDs. The
ideal scaling curve is a horizontal line and none of the libraries are able to
achieve that.  This suggests that there is a significant performance gap that
can be closed, to improve the performance of large-scale DL workloads.

In this paper, we introduce the Performant Collective Communication Library
(PCCL), designed to accelerate collective operations -- specifically
all-gather, reduce-scatter, and all-reduce -- for large buffer sizes ($>$10 MB)
found in parallel deep learning workloads.
Our design exploits hierarchical algorithms, use of CUDA kernels for
reductions, and adaptive selection between different algorithmic choices to provide the
best performance possible.

We evaluate PCCL on Frontier and Perlmutter, dragonfly-based supercomputers
with AMD MI250X and NVIDIA A100 GPUs respectively, demonstrating its 
efficiency and scalability on both
systems.  Our implementations of all-gather, reduce-scatter, and all-reduce
achieve significant speedups over all three libraries -- Cray-MPICH, NCCL, and RCCL.
The optimized collectives in PCCL thus pave the way for scalable performance of
large-scale deep learning workloads on next-generation GPU
supercomputers.

The paper makes the following key contributions:
\begin{itemize}
    \item We analyze the limitations of existing communication libraries --
Cray-MPICH, NCCL, and RCCL on Perlmutter and Frontier for all-gather and
reduce-scatter operations in parallel deep learning workloads.
    \item We develop optimized implementations of all-gather, reduce-scatter,
and all-reduce in PCCL, with a focus on effectively utilizing system resources
and ensuring scalability for large messages and GPU counts.
    \item We demonstrate substantial performance speedups over RCCL on 2048
GCDs of Frontier -- up to 168$\times$ for reduce-scatter, 33$\times$ for
all-gather and 10$\times$ for all-reduce. More modest but still significant
gains up to 5.7$\times$ over NCCL are observed on Perlmutter.
    \item We benchmark the performance of 
multi-billion-parameter LLM training workloads to validate the practical
benefits of our optimizations, demonstrating significant speedups in training
throughput: up to 4.9$\times$ speedup over RCCL in DeepSpeed ZeRO-3 training,
and up to 2.4$\times$ speedup over RCCL in DDP training.
\end{itemize}

\section{Background on Collective Operations}
\label{sec:bg}
In this section, we provide relevant background on the role of collective
communication in distributed DL workloads.

\subsection{Collective Communication in Distributed Deep Learning}

Sharded data parallelism (SDP) and distributed data parallelism (DDP) are
widely used for large scale training~\cite{fsdp, sc2020zero,
yanyou-large-batch}. They are briefly described below.

\begin{figure}[h]
  \centering
  \includegraphics[width=\columnwidth]{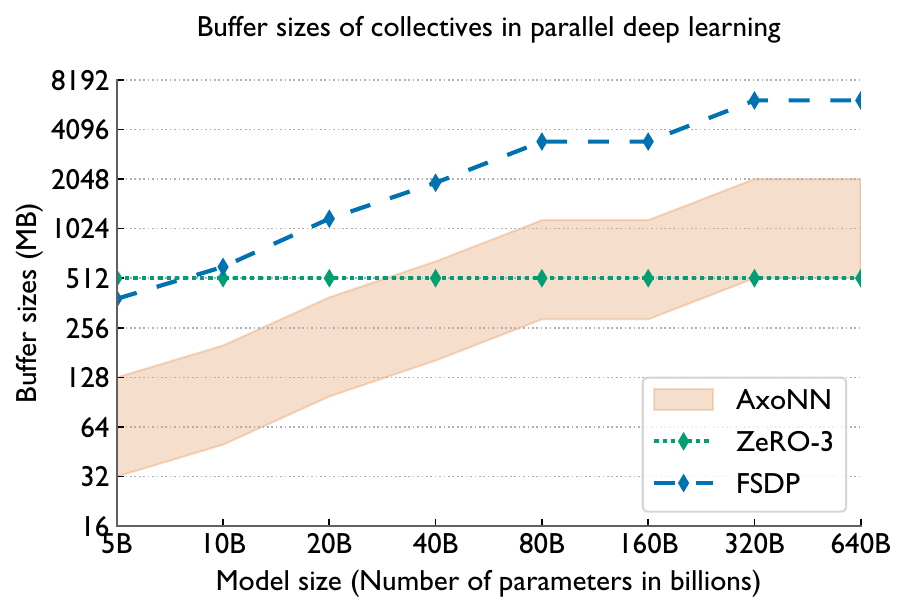}
  \caption{Distribution of all-gather and reduce-scatter message sizes for
several deep learning frameworks for a range of transformer~\cite{transformer} model sizes.
The y-axis represents input buffer sizes for all-gathers but output buffer
sizes for reduce-scatters.}
  \label{fig:msg-sizes}
\end{figure}

\vspace{0.08in}
\noindent{\bf Collective Communication in Sharded Data Parallelism}: In SDP,
Model parameters and gradients are partitioned (or ``sharded'') across GPUs,
which significantly reduces memory requirements and allows for the training of
extremely large models. Two critical collective communication operations --
\textit{all-gather} and \textit{reduce-scatter} -- play a central role. These
operations aggregate distributed data across GPUs: the all-gather operation
collects model parameters from all shards to form a complete copy, while the
reduce-scatter operation performs a reduction and distributes gradients across
participating processes. In Figure~\ref{fig:msg-sizes}, we plot the all-gather
and reduce-scatter message sizes for three frameworks that support sharded data
parallelism -- FSDP~\cite{fsdp}, DeepSpeed ZeRO-3~\cite{sc2020zero}, and
AxoNN~\cite{singh:sc2024}.  Unlike FSDP and ZeRO-3, AxoNN performs all-gathers
and reduce-scatters for each linear layer separately, which results in a wide
range of buffer sizes.  Notice how the message sizes across these three
frameworks are in the tens to hundreds of megabytes, even becoming more than a
gigabyte for larger models.

\vspace{0.08in}
\noindent{\bf Collective Communication in Distributed Data Parallelism}:
In DDP, model parameters are replicated across GPUs, and the \textit{all-reduce} collective is critical for synchronizing parameter replicas.
During training, each GPU performs forward and backward passes on a distinct local mini-batch of the input dataset, producing gradients 
specific to each GPU's mini-batch.
After the backward pass, DDP invokes all-reduce to average the gradients across GPUs, and each GPU performs a local parameter update
to keep all replicas in sync.
Note that the all-reduce message sizes are directly proportional to the number of model parameters. For example, training a 1B-parameter model 
requires an all-reduce on 4 GB of gradients per iteration (one FP-32 gradient scalar per parameter). To reduce latency and improve overlap with computation, PyTorch's DDP framework splits this large all-reduce into several smaller all-reduces with sizes ranging from 48--80 MB \cite{pytorchdist-vldb}, and overlaps them with the backward pass compute.

\begin{figure*}[t]
  \centering
  \includegraphics[width=0.325\textwidth]{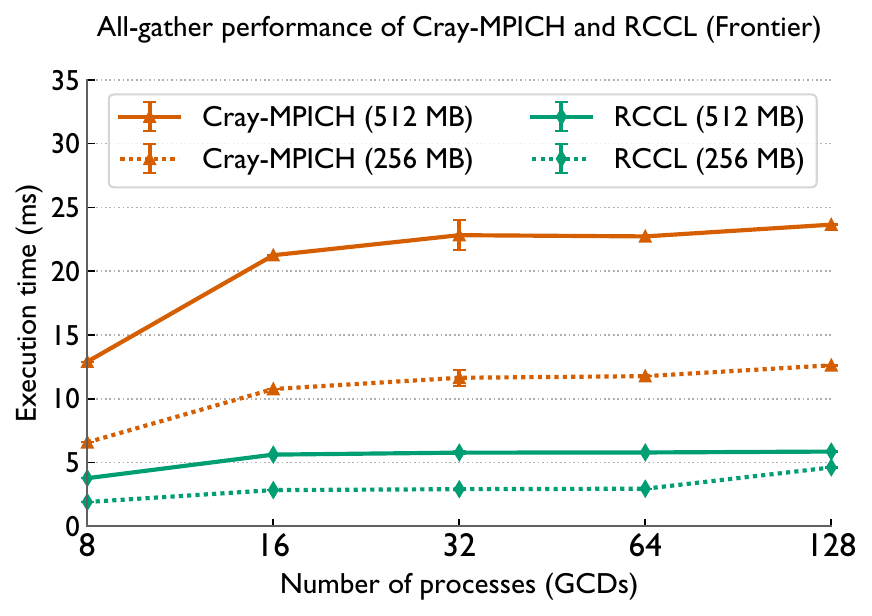}
  \includegraphics[width=0.325\textwidth]{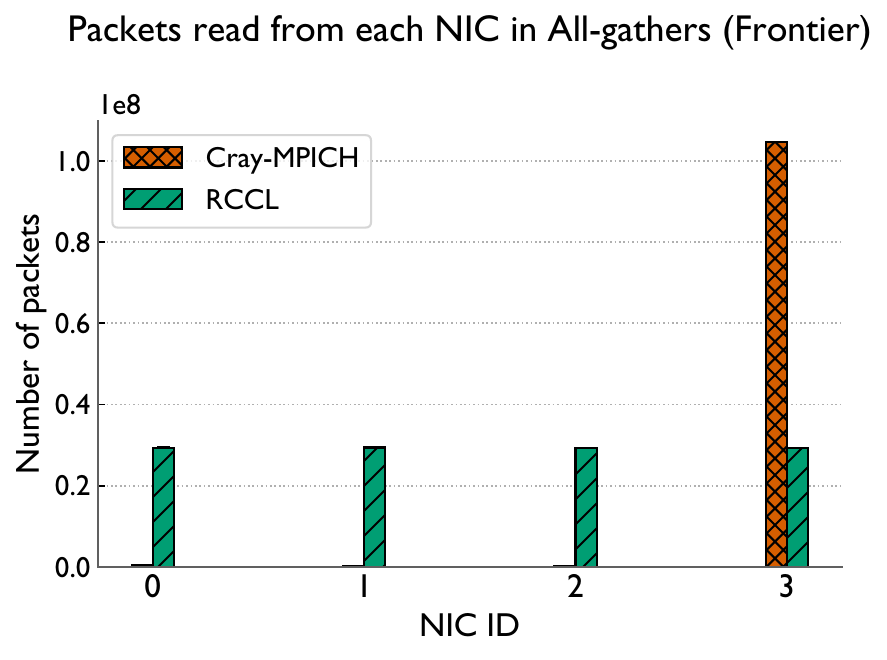}
  \includegraphics[width=0.325\textwidth]{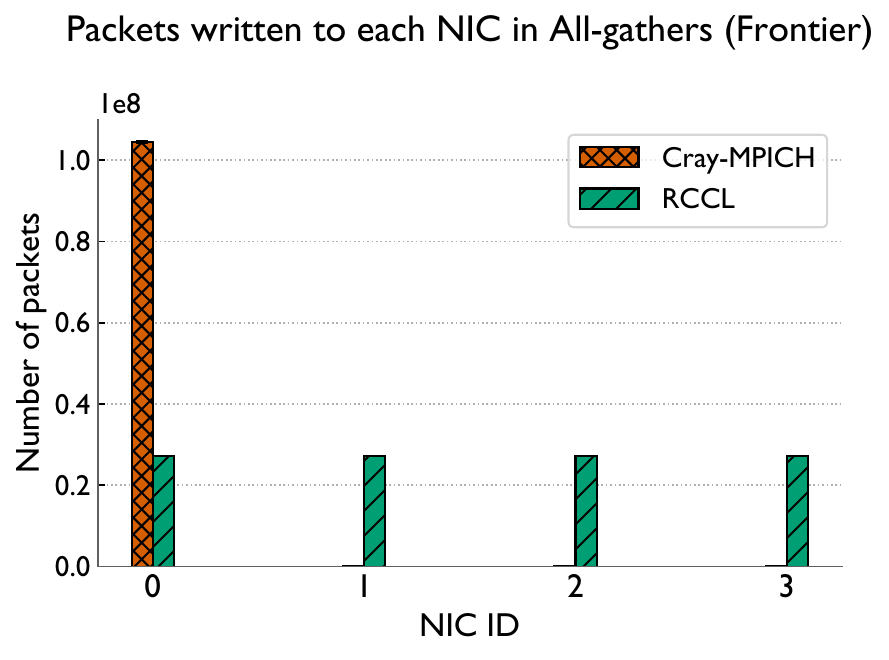}
  \caption{All-gather performance of Cray-MPICH and RCCL
on Frontier for a bandwidth-bound scenario with large message sizes (256 and
512 MB) and small GPU counts (left). The middle and right plots demonstrate the number of
packets read from (middle) and written to (right) each of the four NICs on a
Frontier compute node during all-gather operations.}
  \label{fig:rccl-better}
\end{figure*}

\subsection{Algorithms for Dissolving Different Collectives}
\label{sec:bg-coll}

Efficient implementations of all-gather, reduce-scatter, and all-reduce operations are critical for distributed deep learning. 
In this work, we build on popular algorithms and introduce enhancements to them to improve performance and scalability.

\vspace{0.08in}
\noindent{\bf Ring:}
The ring algorithm is a popular method for implementing collective
communications due to its simplicity and efficiency in certain network
topologies. In a ring-based collective operation, each process communicates
with its immediate neighbors in a circular fashion.  While effective at
moderate scales and large message sizes, the ring algorithm can suffer from
inefficiencies at larger scales due to its latency term being linearly
proportional to the number of processes. The communication time of a ring
all-gather can be modeled as,

    ~\vspace{0.2in}
    \begin{equation}
    T_{\text{ring}} = \eqnmarkbox[color_1]{a}{\alpha} \times (\eqnmarkbox[color_2]{b}{p}-1) + \eqnmarkbox[color_3]{c}{\beta} \times \frac{p-1}{p}\eqnmarkbox[color_4]{d}{m}
    \end{equation}
    \annotate[yshift=.5em]{above,left}{a}{startup latency}
    \annotate[yshift=.5em]{above,right}{b}{number of processes}
    \annotate[yshift=-.25em]{below,left}{c}{inverse of bandwidth}
    \annotate[yshift=-.25em]{below,right}{d}{buffer size}
    \vspace{0.1in}

\noindent where \(p\) is the number of processes, \(\alpha\) represents the startup latency per message, \(m\) is the size of the 
output buffer on each GPU, and \(\beta\) is the inverse of the peer-to-peer bandwidth. 

\vspace{0.08in}
\noindent{\bf Recursive Halving/Doubling:}
A popular way of minimizing latency costs involves utilizing the recursive halving or doubling algorithms for all-gathers and 
reduce-scatters respectively. These algorithms divide the communication task into logarithmically many steps, and hence their performance 
scales better than the ring algorithm. The communication time for a recursive-doubling all-gather can be modeled as,
    \vspace{0.2in}
    \begin{equation}
    T_{\text{rec}} = \eqnmarkbox[color_1]{a}{\alpha} \times \log_2(\eqnmarkbox[color_2]{b}{p}) + \eqnmarkbox[color_3]{c}{\beta} \times \frac{p-1}{p}\eqnmarkbox[color_4]{d}{m}
    \end{equation}
    \annotate[yshift=.5em]{above,left}{a}{startup latency}
    \annotate[yshift=.5em]{above,right}{b}{number of processes}
    \annotate[yshift=-.25em]{below,left}{c}{inverse of bandwidth}
    \annotate[yshift=-.25em]{below,right}{d}{buffer size}
    \vspace{0.1in}

For small message sizes or very large process counts, the logarithmic growth in the latency term often leads to lower overall communication costs. 
Building on these primitives, an all-reduce can be implemented as a reduce-scatter followed by an all-gather operation.
More details about these and other algorithms can be found in Thakur et al.~\cite{thakurimproving2003}.

\section{Current State of Communication Libraries}
\label{sec:issues}

We begin with investigating the current state of popular communication libraries
-- Cray-MPICH, NCCL, and RCCL, on Perlmutter and Frontier. We find
unique issues that limit the performance of each library, and highlight
these below.

\subsection{Benchmarking Setup}
\label{sec:issue-methodology}

To ensure a fair comparison, we follow best practices for benchmarking all three
libraries -- optimal process placement, NUMA-aware NIC binding, disabling eager
messaging, and enabling GPU Direct RDMA~\cite{sensiexploring2024}.
 

\vspace{0.08in}
\noindent{\bf Message Sizes:}
In line with Figure~\ref{fig:msg-sizes}, our evaluation in this section focuses
on two message sizes -- 256 MB and 512 MB. Note that for all-gathers and
reduce-scatters, these values refer to the output and input message size per
GPU respectively. 


\vspace{0.08in}
\noindent{\bf Software Stack:}
On Frontier, our software stack comprises of ROCm 6.2.4, RCCL 2.20.5,
Cray-MPICH 8.1.31, libfabric 1.15.2 and the aws-ofi-rccl plugin
version v1.4. On Perlmutter, we use CUDA 12.4, NCCL 2.24.3, Cray-MPICH
8.1.30, and libfabric 1.22.0.

\vspace{0.08in}
For each combination of library, collective, message size, and GPU count, we
perform ten independent trials. We measure the total time spent in the
collective during each run using AMD's \texttt{hipeventtimers} instrumentation
on Frontier, and NVIDIA's \texttt{cudaEventElapsedTime} on Perlmutter. All
reported timings are end-to-end measurements that include any necessary data
movement and transformation costs, including intra-GPU transposes required by
hierarchical algorithms. We compute the mean and standard deviation over the
ten trials to ensure statistical robustness. 

\subsection{Resource Under-utilization in Cray-MPICH}
\label{sec:issue-nic-underutil}


Figure~\ref{fig:rccl-better} (left) presents a comparative analysis of all-gather performance between Cray-MPICH and RCCL on
Frontier for message sizes of 256 MB and 512 MB. We observe that RCCL achieves approximately a $4\times$ performance advantage in this
bandwidth-bound scenario. To explain this disparity, we examine two hardware performance counters provided by the
Cassini Slingshot-11 Network Interface Controllers (NICs)~\cite{cassini_nic_counters} -  \texttt{\seqsplit{parbs\_tarb\_pi\_posted\_pkts}} and
\texttt{\seqsplit{parbs\_tarb\_pi\_non\_posted\_pkts}}.
These counters represent the count of packets read from and written to each NIC within a node during job execution, respectively. As shown in
Figure~\ref{fig:rccl-better} (middle, right), we observe a consistent trend across all runs -- Cray-MPICH exclusively utilizes NIC-0 for all write operations, and NIC-3 for all read
operations. Conversely, RCCL distributes its read and write operations evenly across all four NICs on the node. This
underutilization of available network resources by Cray-MPICH explains the $4\times$ performance gap we observe in all-gather
operations. 

\begin{figure}[h]
  \centering
  \includegraphics[width=\columnwidth]{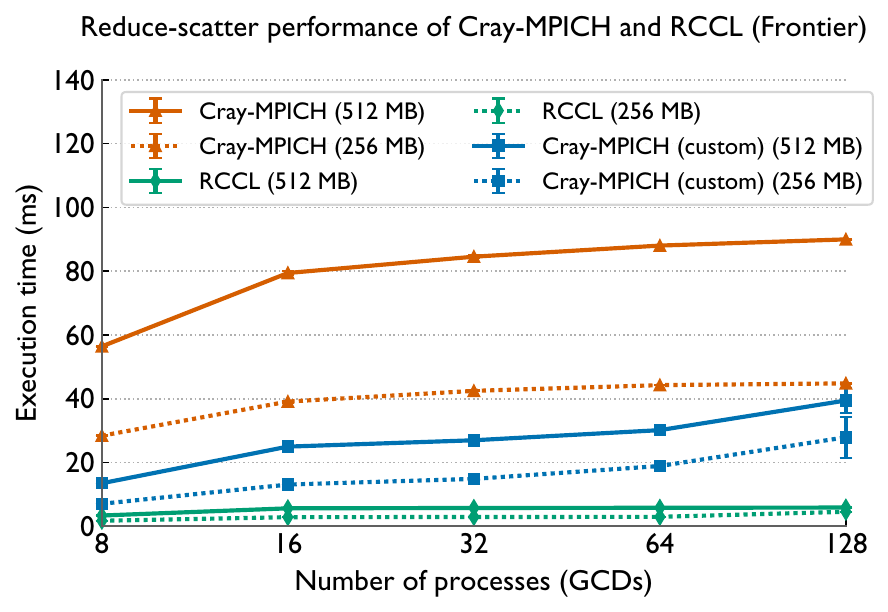}
  \caption{Performance comparison of reduce-scatter using Cray-MPICH, RCCL, and
a custom implementation of reduce-scatter that uses MPI point-to-point primitives and GPU
compute kernels.}
  \label{fig:reduce-scatter-cpu}
\end{figure}

Moreover, for collectives that require computation, such as reduce-scatters and all-reduces, Cray-MPICH suffers
from another performance bottleneck -- it performs
the reduction operations on the CPU instead of offloading them to the GPU. This introduces significant computational overheads, as shown in
Figure~\ref{fig:reduce-scatter-cpu}. Cray-MPICH (orange) performs significantly
worse than RCCL (green). Notably, this performance gap is far greater than the
$4\times$ difference observed earlier for all-gather in
Figure~\ref{fig:rccl-better}.

\observe{issue-cray-mpich}{Cray-MPICH severely underutilizes available network (NIC) and computational (GPU) resources.
It routes all network traffic through a single NIC, and performs reduction operations on the CPU, instead of offloading
them to the GPU.}

While the NIC underutilization issue outlined previously
still persists for Cray-MPICH reduce-scatter operations, it alone cannot explain this performance disparity.
We hypothesize that this disparity stems from Cray-MPICH's choice to perform reductions on the CPU instead
of the more powerful GPUs. To validate this, we manually implemented the reduce-scatter operation using
MPI point-to-point primitives and a GPU vector reduction kernel. As shown in
Figure~\ref{fig:reduce-scatter-cpu}, our implementation (blue line) achieves
performance that is several times faster than Cray-MPICH's native
reduce-scatter, further supporting our hypothesis.







\subsection{Poor Scaling of NCCL and RCCL at Large GPU Counts}


Performance limitations of Cray-MPICH have made NCCL and RCCL popular for distributed
deep learning, but as we demonstrate next, these libraries have algorithmic
limitations that prevent them from scaling efficiently to large GPU counts.
Let us again look at all-gather performance of RCCL (Frontier) and
NCCL (Perlmutter) in Figure~\ref{fig:high-latencies}. We observe that both of
these libraries scale extremely poorly. On investigating deeper, we found that
NCCL and RCCL only support the ring algorithm for all-gather and
reduce-scatter (refer to Section~\ref{sec:bg-coll}).  While effective for
bandwidth-bound workloads, the ring algorithm performs poorly in
latency-sensitive scenarios because each process must send and receive $(p-1)$
messages sequentially, causing the total communication time to grow linearly
with the number of processes.

Notably, while NCCL and RCCL implement a logarithmic latency scaling
algorithm for all-reduce based on the double-binary tree
structure~\cite{hu2025demystifyingncclindepthanalysis}, for all-gather
and reduce-scatter, they lack log latency scaling algorithms such as recursive doubling or halving (refer to Section~\ref{sec:bg-coll}). These are known to reduce the number of communication steps to $\log_2 p$ and
are generally preferred for small message sizes or high process counts. 
NCCL has recently introduced a logarithmic scaling algorithm for all-gather and reduce-scatter called PAT. However, at the time of writing, it only supports one GPU per node~\cite{jeaugey2025patnewalgorithmallgather}.
This
lack of algorithmic diversity directly contributes to the sub-optimal scaling
we observe at large GPU counts.

\observe{exp:scaling}{NCCL and RCCL rely solely on the ring algorithm for all-gather and reduce-scatter, leading to poor 
scaling in latency-bound scenarios. More efficient algorithms such as recursive doubling and halving are not supported.}

\section{Design of Scalable Collectives in PCCL}
\label{sec:solutions}
\begin{figure*}[t]
\centering
\includegraphics[width=\textwidth]{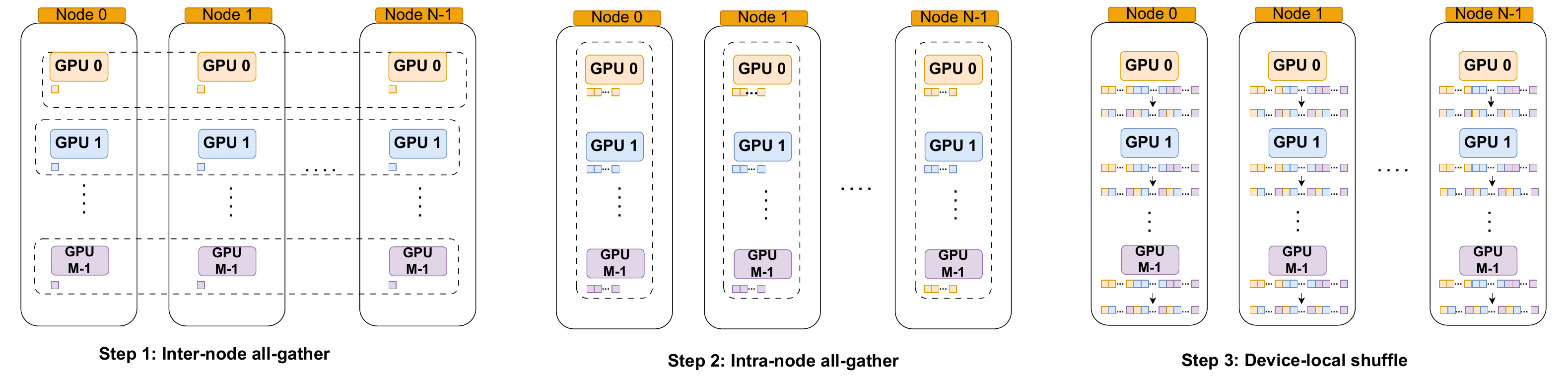}
\caption{Diagram showing our hierarchical (two-level) implementation to
dissolve an all-gather operation on a GPU-based cluster with N nodes and M GPUs
per node. In step 1, we perform inter-node all-gathers, in step 2, we perform
intra-node all-gathers, and in step 3, each GPU performs a local shuffle of the
received data.}
\label{fig:schem}
\end{figure*}

In Section~\ref{sec:issues}, we identified several challenges affecting
Cray-MPICH, NCCL, and RCCL, in the context of collective communication for deep
learning workloads. These challenges create significant barriers to efficiently
scaling DL workloads across thousands of GPUs. We now present the Performant
Collective Communication Library (PCCL), which addresses these challenges
through a three-pronged approach. 

First, PCCL provides the option to directly call existing collective libraries
(Cray-MPICH, NCCL, or RCCL) when they perform well for a given message size and
GPU count. Second, PCCL provides new latency-optimized implementations of
collectives for scenarios where existing libraries are sub-par -- specifically,
training at large-scale, as identified in Section~\ref{sec:issues}. These new
implementations employ a hierarchical intra-node and inter-node design with two
inter-node algorithm backends: \texttt{PCCL\_ring} (ring-based) and
\texttt{PCCL\_rec} (recursive doubling/halving). Third, PCCL includes a
learning-based adaptive dispatcher that automatically selects the most
performant option at runtime based on the specific workload characteristics.
The dispatcher can choose from any of the available backends: the existing
libraries (Cray-MPICH, NCCL, RCCL) or the new proposed hierarchical collectives
-- (\texttt{PCCL\_ring} and \texttt{PCCL\_rec}). This design allows PCCL to
achieve strong performance across the full spectrum of configurations—from
small-scale runs with large messages to large-scale runs with small messages.
Below, we discuss the design of our new latency-optimized collective
implementations, followed by a detailed description of the learning-based
adaptive dispatcher.

\subsection{Hierarchical Collective Algorithms for Scalability}

Our optimized implementations of all-gather, reduce-scatter, and all-reduce are
based on a two-level hierarchical design. While our primary motivation is to
address the underutilization of NICs (identified in
Section~\ref{sec:issue-nic-underutil}), this design also reduces latency and
improves scalability~\cite{choblueconnect2019, nguyen2hierarchical2018}.

We illustrate our design in Figure~\ref{fig:schem} for an all-gather operation
on a hypothetical system with $N$ nodes and $M$ GPUs per node. The global
collective operation to be performed across all GPUs is divided into two
distinct phases (inter-node and intra-node) using sub-communicators. Inter-node
sub-communicators are formed by grouping together corresponding GPUs (with the
same local ID) across nodes. For example, in Figure~\ref{fig:schem}, all GPUs
with the same within-node ID are grouped together to create a total of $M$
inter-node sub-communicators. Similarly, intra-node sub-communicators are
formed by grouping together all GPUs within a node to create $N$ intra-node
sub-communicators.

The hierarchical communication for dissolving the collective unfolds in three
steps. First, we schedule concurrent all-gather operations within all
inter-node sub-communicators (Step 1 of Figure~\ref{fig:schem}). After the
completion of this phase, each GPU has received data from its counterparts in every other node,
yielding a partial result distributed across GPUs within each node. Second,
we perform an intra-node all-gather (Step 2 of Figure~\ref{fig:schem}),
after which each GPU holds the complete output in memory, albeit in an
incorrect order. Third, a device-local shuffle (Step 3 of
Figure~\ref{fig:schem}) rearranges each GPU's data into the correct order;
in practice, this is implemented as a transpose kernel. We implement
reduce-scatter similarly, starting with the intra-node phase followed by the
inter-node phase. All-reduce is implemented by composing a two-level
reduce-scatter with a two-level all-gather.

This design addresses NIC under-utilization by scheduling all inter-node
all-gathers in Step 1 concurrently.  On Frontier, each node has four NICs
connected to two GCDs each. We ensure each GCD exclusively uses its
corresponding NIC (e.g., GCDs 0 and 1 use NIC 0, GCDs 2 and 3 use NIC 1, etc.),
evenly distributing inter-node traffic across all NICs.
While our hierarchical design explicitly exploits the intra-node topology by
mapping GPUs to their respective NICs, we do not explicitly map our
communication patterns to the inter-node network topology. On Dragonfly-based
systems such as Frontier and Perlmutter, UGAL routing~\cite{ugal} distributes inter-node
traffic across the network, reducing the need for topology-aware collective
mapping.

\vspace{0.05in}
\noindent{\bf Choice of Communication Libraries:}
Our choice of libraries for each level of the hierarchy is driven by
performance and reliability.
For inter-node communication, we
use MPI due to RCCL failures at scale, as reported in prior
work~\cite{geiping2025scalingtesttimecomputelatent} and noted by
HPE\footnote{\url{https://www.olcf.ornl.gov/wp-content/uploads/OLCF_AI_Training_0417_2024.pdf}}
and the OLCF User
Guide\footnote{\url{https://docs.olcf.ornl.gov/software/analytics/pytorch_frontier.html\#environment-variables}}.
Moreover, we find that Cray-MPICH exhibits lower performance variability
than RCCL on Slingshot interconnects, further motivating its use at the
inter-node level.

For intra-node communication, we use GPU-vendor
libraries (NCCL or RCCL) as they are highly optimized for intra-node
communication, efficiently utilizing shared memory, PCIe, and Infinity Fabric or
NVLink connections~\cite{sensiexploring2024}.  NCCL and RCCL only support the
ring algorithm for intra-node all-gather and reduce-scatter.  Since the ring
algorithm is well-suited when the number of GCDs/GPUs per node is small
(eight on Frontier, and four on Perlmutter), we adopt this for all intra-node
collectives.  This ensures that the GPU-GPU link bandwidth is saturated
effectively.

\vspace{0.08in}
\noindent{\bf Messaging Protocol Selection:}
For intra-node communication, we rely on the vendor libraries' heuristics for internal
protocol selection, such as NCCL and RCCL's ability to switch to
low-latency (LL/LL128) protocols for small messages on NVLink and Infinity
Fabric. While Cray-MPICH does not support these low-latency protocols for
inter-node communication, this is not a concern as the scope of our work is
focused on large message sizes (16 MB to 1 GB).

\subsection{Custom Implementations for Inter-node Operations}



The use of MPI for collective operations in the inter-node phase presents
some challenges. With potentially thousands of GPUs participating in the
collective, the choice of the communication algorithm becomes critical for
performance. However, Cray-MPICH only offers the ring algorithm.  As described
in Section~\ref{sec:bg-coll}, the ring algorithm's linear scaling in latency
with respect to the number of processes makes it sub-optimal at large-scale.
Additionally, as discussed in Section~\ref{sec:issue-nic-underutil}, CPU-based
compute in Cray-MPICH limits performance.
Hence, we have developed custom implementations of each collective in
PCCL for the inter-node phase that use MPI point-to-point send and receive
messages. We implement two different algorithms as described below.

\vspace{0.08in}
\noindent{\bf Ring Algorithm:} We implement an inter-node version of the ring
algorithm described in Section~\ref{sec:bg-coll}.  This can be more performant
in bandwidth-bound scenarios. We refer to our implementation of the ring
algorithm as \texttt{PCCL\_ring}. For reduce-scatter and all-reduce, we schedule the local reduction computation on 
the GPU to improve performance, as opposed to performing it on the CPU as in Cray-MPICH.

\vspace{0.08in}
\noindent{\bf Recursive Doubling/Halving Algorithm:} We utilize the recursive
doubling algorithm for inter-node all-gather and recursive halving for
inter-node reduce-scatter operations~\cite{thakurimproving2003}, and implement
them using MPI point-to-point messages.  These algorithms offer logarithmic
latency terms (refer to Section~\ref{sec:bg-coll}), enabling significantly better scaling
performance as the number of GPUs increases. Similar to our ring implementation for reduce-scatter operations, we also ensure that our local vector reduction computation is efficient by
scheduling it on GPU cores. We refer to our implementation using recursive
doubling/halving as \texttt{PCCL\_rec}. Further, our all-reduce in \texttt{PCCL\_rec} uses recursive halving followed
by recursive doubling for inter-node traffic.

\begin{figure}[h]
\centering
\includegraphics[width=\columnwidth]{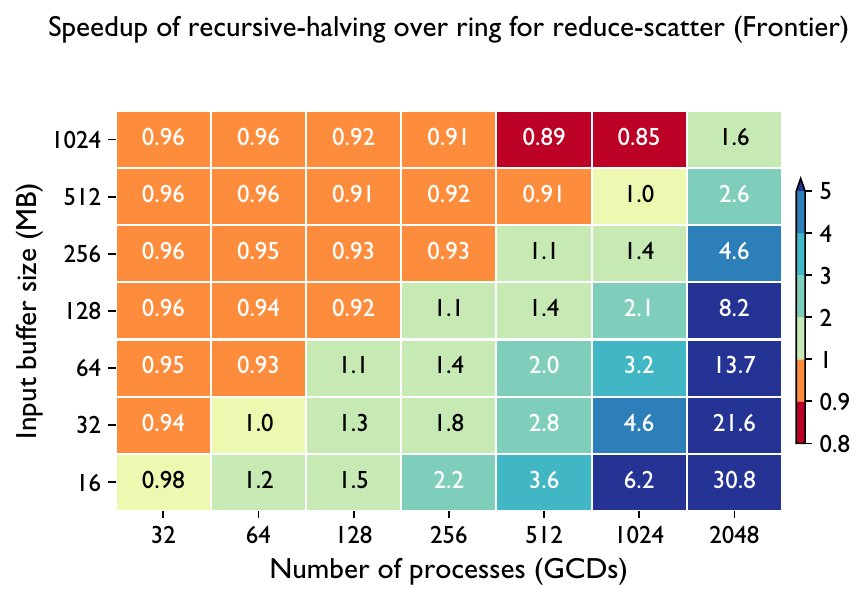}
\caption{Heatmap showing speedups from using recursive halving over
the ring algorithm in the inter-node phase of the reduce-scatter implementation in PCCL.
}
  \label{fig:comm-alg-speedup}
\end{figure}

Figure~\ref{fig:comm-alg-speedup} shows the speedup of recursive halving over
ring for inter-node reduce-scatter.  In bandwidth-bound scenarios (fewer
processes and/or larger messages), ring performs better.  In latency-bound
scenarios (more processes and/or smaller messages), recursive halving achieves
significant improvements, confirming our algorithmic complexity analysis. Our
goal is to be able to choose the ideal algorithm in different scenarios, and we
describe our efforts in that direction below.

We have developed these implementations in C++ as part of PCCL and expose Pybind11
bindings to enable seamless integration with Python-based deep learning
frameworks such as ZeRO-3~\cite{sc2020zero}. Implementing these algorithms in
C++ proves to be critical for achieving high performance.


\subsection{Learning-based Adaptive Dispatching}
\label{sec:adaptive-dispatching}

\begin{figure}[h]
\centering
\includegraphics[width=\columnwidth]{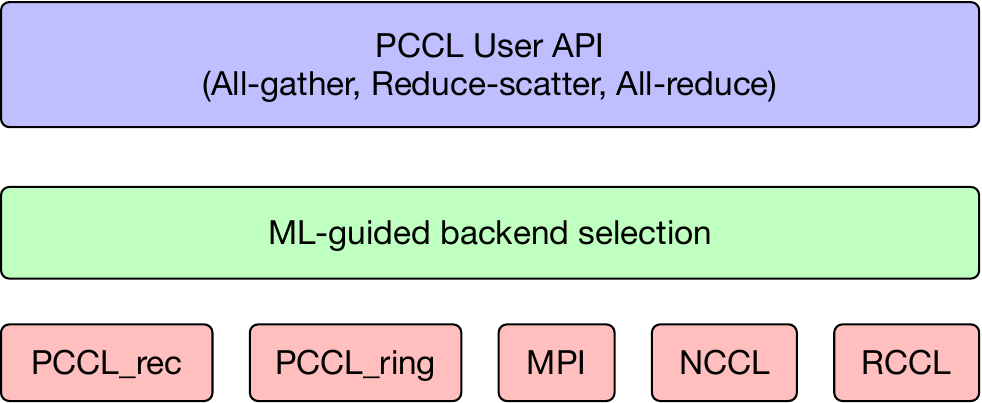}
\caption{The ML-guided selection mechanism in PCCL can enable choosing the best performing backend from several available options.}
  \label{fig:pccl-schematic}
\end{figure}

We have observed, based on our empirical analysis, that no single communication
library or algorithm -- Cray-MPICH, NCCL, RCCL, \texttt{PCCL\_ring}, or \texttt{PCCL\_rec}
-- is universally fastest. Performance depends on the specific configuration, with
message size and GPU count being the dominant factors. For example, RCCL often
outperforms PCCL’s ring/recursive algorithms at lower GPU counts and larger
message sizes.  This motivated us to develop an adaptive dispatcher capable of
selecting the most suitable library at runtime. This modified design of the
PCCL library is shown in Figure~\ref{fig:pccl-schematic}. 

In order to enable optimal backend selection, we implemented a lightweight
dispatcher based on Support Vector Machines (SVMs)~\cite{Cortes:1995,
Smola:2004} -- a supervised learning algorithm that classifies data by finding
the optimal decision boundary that separates distinct categories. SVMs work by
identifying the hyperplane that maximizes the margin between different classes
in the feature space, allowing for effective classification even in complex
scenarios.  For each machine and collective pair, we train a dedicated SVM
classifier using empirical data spanning message sizes from 1 MB to 1024 MB and
GPU counts from 4 to 2048.  For each configuration (library, message size, GPU
count), we include ten independent runs in our dataset. This dataset is
partitioned using a stratified 80/20 train-test split to maintain class
balance. Hyperparameter selection for each SVM is performed via five-fold
cross-validation on the training set, ensuring robust model selection and
mitigating overfitting. At runtime, the dispatcher queries the appropriate trained SVM
with the GPU count and message size as input features to predict the optimal
backend. We evaluate the performance of the trained model on 20\% unseen test
data. As Table~\ref{tab:svm_accuracy} shows, the high prediction accuracy and
low misclassification rates indicate that the dispatcher generalizes well to
previously unseen configurations.

\begin{table}[h]
    \centering
    \caption{SVM dispatcher performance on the unseen test set (20\% of data).}
    \label{tab:svm_accuracy}
    \begin{tabular}{@{}llrrr@{}}
      \toprule
      \textbf{Machine} & \textbf{Collective} & \makecell{\textbf{Test} \\ \textbf{Size}} & \makecell{\textbf{Correctly} \\ \textbf{Classified}} & \textbf{Accuracy (\%)} \\
      \midrule
      \multirow{3}{*}{Frontier}
      & All-gather      & 20 & 17 & 85.0 \\ 
      & Reduce-scatter  & 20 & 18 & 90.0 \\ 
      & All-reduce      & 20 & 16 & 80.0 \\ 
      \midrule
      \multirow{3}{*}{Perlmutter}
      & All-gather      & 22 & 20 & 90.9 \\ 
      & Reduce-scatter  & 22 & 21 & 95.4 \\ 
      & All-reduce      & 20 & 15 & 75.0 \\ 
      \bottomrule
    \end{tabular}
\end{table}


\section{Experimental Setup}
\label{sec:setup}
Next, we provide details of the experimental setup for our comparison of
PCCL collectives with other communication libraries, and in the context
of production DL workloads.

\subsection{Benchmarking Collective Operations}
\label{sec:setup-raw-collectives}

First, we compare the standalone performance of all-gather, reduce-scatter, and
all-reduce operations, which are the primary focus of this paper.  Our
experiments cover a range of message sizes from 16 MB to 1 GB. For all-gather,
this range denotes each GPU's output buffer size; for reduce-scatter, the
input buffer size; for all-reduce, both input and output buffer sizes. For
each message size, we measure performance across 32 to 2048 GCDs (4 to 256
nodes) on Frontier, and 32 to 2048 GPUs (8 to 512 nodes) on Perlmutter.

We conduct ten independent trials for each combination of library, collective,
message size, and GPU count.  We consider the following libraries on each
system: Cray-MPICH (the default MPI implementation on HPE Cray systems with Slingshot interconnects), NCCL or RCCL, and PCCL.  For both systems, our setup
for process placement, measurement protocol, communication tuning, and software
stack is consistent with the configuration described in
Section~\ref{sec:issue-methodology}. For all-reduce, at the time of our runs, a
more recent software stack was available: on Frontier we use ROCm
6.4.1, RCCL 2.22.3, Cray MPICH 8.1.32,
libfabric 1.22.0 and the aws-ofi-rccl plugin version v1.4. On
Perlmutter, we use CUDA 12.9, NCCL 2.27.3, 
Cray-MPICH 8.1.30, and libfabric 1.22.0.

\subsection{Comparative Evaluation of Production DL Workloads}
\label{sec:setup-e2e}

We use two deep learning workloads to study the impact of using improved implementations in PCCL on overall application performance.

\vspace{0.08in}
\noindent{\bf DeepSpeed ZeRO-3:}
To evaluate the practical benefits of PCCL's all-gather and reduce-scatter, we measure the end-to-end training performance of large language models using DeepSpeed 
ZeRO-3~\cite{sc2020zero}, a widely adopted parallel deep learning framework. We perform strong scaling experiments on seven billion and 13 billion parameter 
GPT-style transformer models~\cite{gpt-3} using model hyperparameters from Zhang et al.~\cite{opt:zhang2022opt}. We list these hyperparameters in 
Table~\ref{tab:setup-perf-gpt}. We use a global batch size of four million tokens and a sequence length of 2048 and use the 
OpenWebText~\cite{openwebtext} corpus to create our training data.

\begin{table}[h]
  \centering
  \caption{\label{tab:setup-perf-gpt} Architectural details of the GPT-style transformer models~\cite{gpt-3} used in the experiments.
  We borrow these hyperparameters from Zhang et al.~\cite{opt:zhang2022opt}.}
  \begin{tabular}{lccccc}
    \toprule
    \textbf{Model} & \textbf{Framework} & \textbf{Params} & \textbf{Layers} & \makecell{\textbf{Hidden}\\\textbf{Size}} & \textbf{Heads} \\ \midrule
    GPT-7B      &   ZeRO-3    & 7B        & 32        & 4096        & 32        \\
    GPT-13B     &   ZeRO-3    & 13B       & 40        & 5120        & 40        \\ 
    GPT-1.3B    &   DDP       & 1.3B      & 24        & 2048        & 32        \\ \bottomrule
  \end{tabular}
\end{table}

\vspace{0.08in}
\noindent{\bf PyTorch DDP:}
PyTorch DDP is a distributed data parallel framework which relies on all-reduce (unlike ZeRO-3). To evaluate the practical benefits of PCCL's all-reduce, we use PyTorch DDP~\cite{pytorchdist-vldb} to train a
1.3 billion parameter GPT-style transformer model~\cite{gpt-3}, with hyperparameters summarized in Table~\ref{tab:setup-perf-gpt}. We use a global batch size of one million 
tokens and use the OpenWebText~\cite{openwebtext} corpus as training data. 

For both ZeRO-3 and DDP experiments, on Frontier, we scale from 128 to 1024 GCDs, and on Perlmutter, we scale from 256 to 2048 GPUs. We run experiments with RCCL 
(Frontier) and NCCL (Perlmutter) as baselines, then swap in \texttt{PCCL\_rec} to handle all-gather and reduce-scatter in ZeRO-3, and to handle all-reduce in DDP. For each 
configuration, we run 10 training batches across three trials and compute the average time per iteration over the last eight batches in each run to minimize warm-up effects.  
We observed significant variability in RCCL all-reduce performance; to account for this, we ran five trials for DDP experiments and report results from the three trials with the smallest mean batch time.

\begin{figure*}[t]
  \centering
  \includegraphics[width=0.325\textwidth]{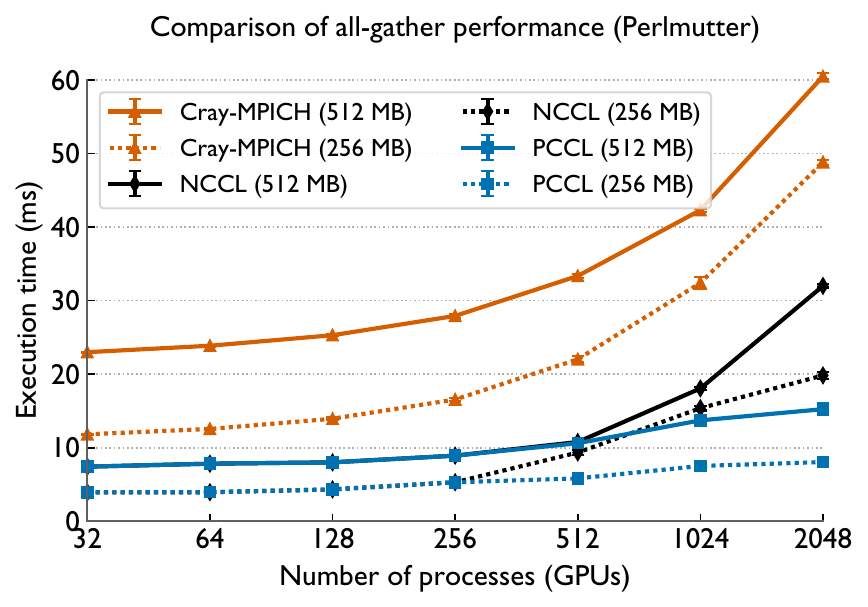}
  \includegraphics[width=0.325\textwidth]{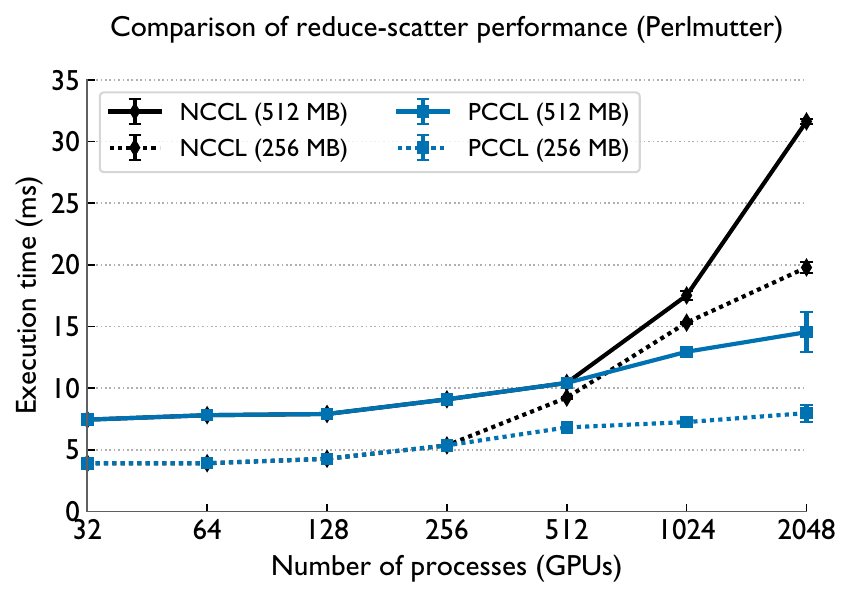}
  \includegraphics[width=0.325\textwidth]{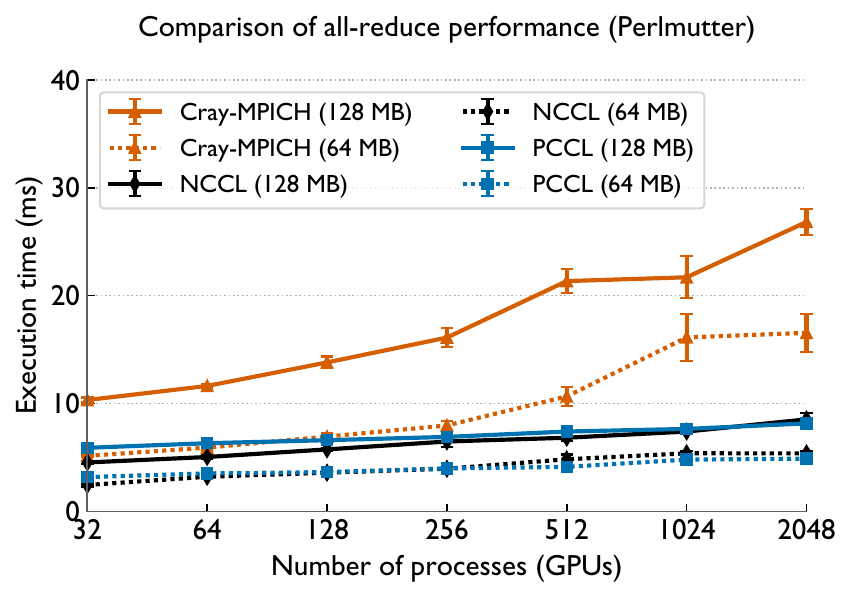}
  \caption{Performance comparison of all-gather (left), reduce-scatter (middle), and all-reduce (right) using Cray-MPICH, NCCL, and
PCCL with adaptive dispatching, for different per-process buffer sizes (256 and 512 MB for all-gather and reduce-scatter; 64 and 128 MB for all-reduce) and varying process counts on Perlmutter.}
  \label{fig:compare-libs-perlmutter}
\end{figure*}

\section{Performance Results}
\label{sec:results}
\begin{figure*}[t]
  \centering
  \includegraphics[width=0.325\textwidth]{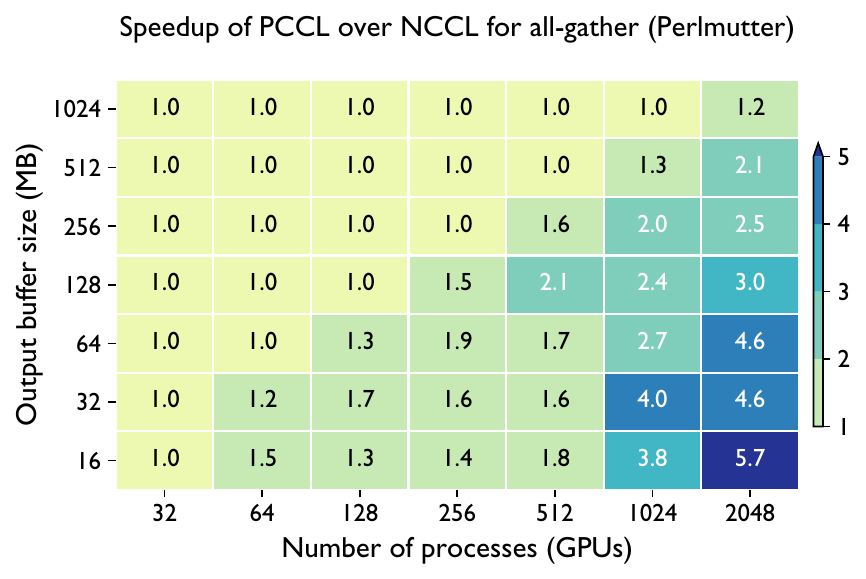}
  \includegraphics[width=0.325\textwidth]{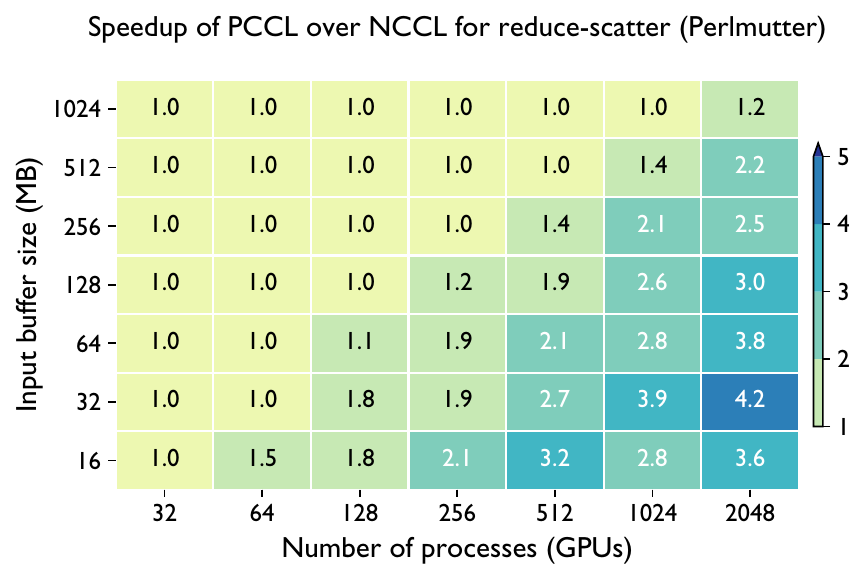}
  \includegraphics[width=0.325\textwidth]{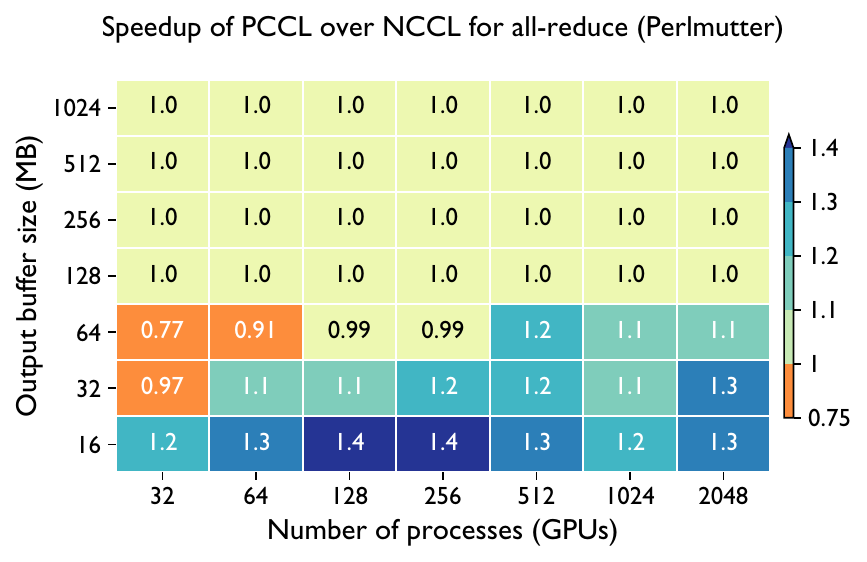}
  \caption{Heatmaps showing speedups from using PCCL with adaptive dispatching over NCCL for all-gather
(left), reduce-scatter (middle), and all-reduce (right) on Perlmutter.  The speedup is shown as a
function of per-process output/input buffer size (in MB) and process count.}
  \label{fig:heatmap-perlmutter}
\end{figure*}

We now present performance comparisons of PCCL with other collective libraries
using benchmarks and in the context of DL applications.

\subsection{Comparisons with Cray-MPICH and NCCL on Perlmutter}

Figure~\ref{fig:compare-libs-perlmutter} presents results for all-gather (left)
and reduce-scatter (middle) collectives on Perlmutter for representative
per-process message sizes of 256 and 512 MB. For all-gather and reduce-scatter,
we observed that both Cray-MPICH (orange lines) and NCCL (black lines) fell short of
ideal scaling, which should be a flat horizontal line.  NCCL's performance
begins to degrade noticeably beyond 512 processes. In contrast, PCCL scales
nearly perfectly across both collectives, maintaining desirable performance
curves, and achieving speedups in the range of $1.3$ -- $4.6\times$ over NCCL
and $8.8$--$15\times$ over Cray-MPICH on 1024 and 2048 GPUs.
Figure~\ref{fig:compare-libs-perlmutter} (right) presents results for
all-reduce operations with message sizes of 64 MB and 128 MB.  Both NCCL and
PCCL exhibit strong scalability with node count. NCCL employs double-binary
trees to achieve log-latency scaling, which explains why the performance of
NCCL and PCCL is nearly identical for the all-reduce operation.

\begin{figure*}[t]
  \centering
  \includegraphics[width=0.325\textwidth]{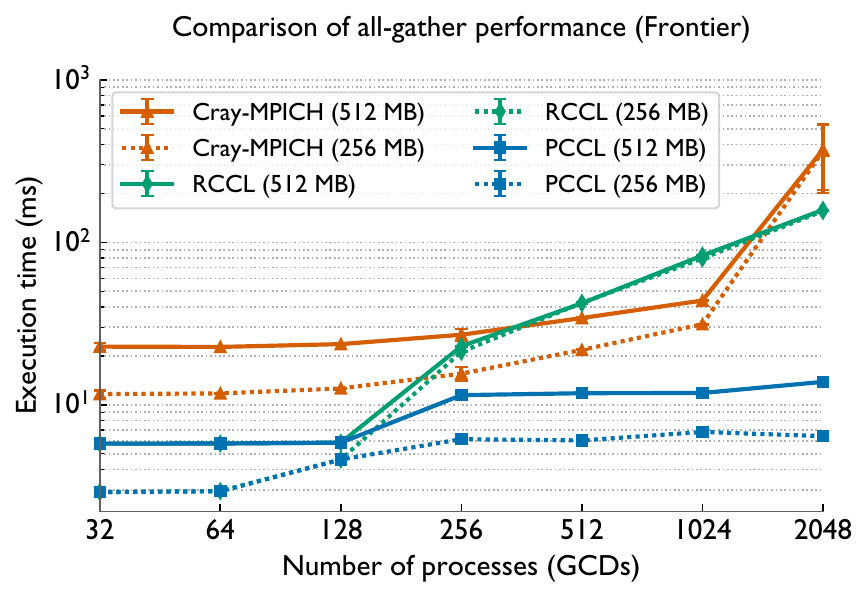}
  \includegraphics[width=0.325\textwidth]{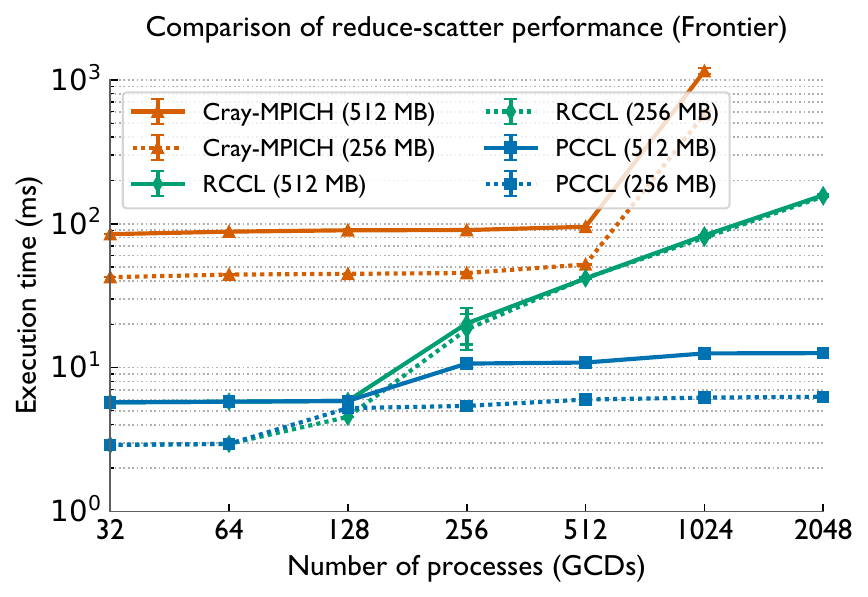}
  \includegraphics[width=0.325\textwidth]{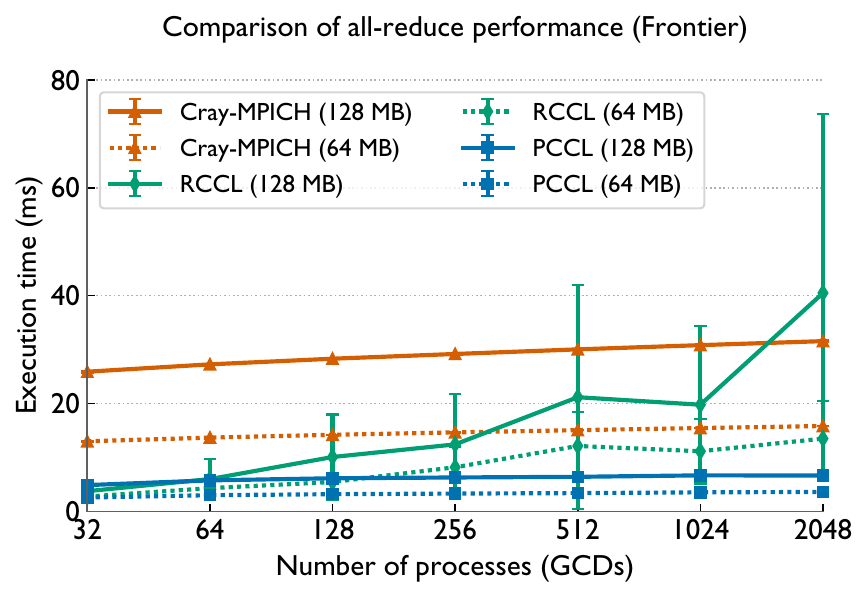}
  \caption{Performance comparison of all-gather (left), reduce-scatter (middle), and all-reduce (right) using Cray-MPICH, RCCL, and
PCCL with adaptive dispatching, for different per-process buffer sizes (256 and 512 MB for all-gather and reduce-scatter; 64 and 128 MB for all-reduce) and varying process counts on Frontier.}
  \label{fig:compare-libs-frontier}
\end{figure*}

Figure~\ref{fig:heatmap-perlmutter} presents heatmaps that present the relative
performance improvement of PCCL over NCCL for a larger range of message sizes
and GPU counts. For all-gather (left) and reduce-scatter (middle), PCCL
performs similar to NCCL in the top-left regions of the heatmaps, representing
bandwidth-bound scenarios. As expected, our adaptive dispatching protocol
selects NCCL for these cells.
However, as we transition to latency-sensitive regions in the bottom-right corners
of the heatmap, PCCL's advantages become evident. Around 1024--2048 processes
and 16--32MB message sizes, PCCL achieves significant speedups over NCCL,
ranging from $3$--$5\times$. While speedups for larger message sizes are
smaller, they remain notable. For example, at 2048 processes and 128--512 MB
message sizes, PCCL is approximately $2$--$3\times$ faster than NCCL, which is still a significant improvement. For all-reduce (right), we observed similar performance between
PCCL and NCCL as both libraries use log-latency scaling algorithms.

\subsection{Comparisons with Cray-MPICH and RCCL on Frontier}

Next, we evaluate PCCL's effectiveness on Frontier, which features AMD MI250X
GPUs.  Figure~\ref{fig:compare-libs-frontier} (left) shows the performance of
all-gather operations on Frontier using PCCL and other communication libraries
for output buffer sizes 256 and 512 MB.  For each configuration, we scale the
number of GCDs from 32 to 2048. Since the output buffer size per GPU remains
fixed, the ideal performance curve for each buffer size is a flat horizontal
line, indicating perfect scaling.

\begin{figure*}[t]
  \centering
  \includegraphics[width=0.325\textwidth]{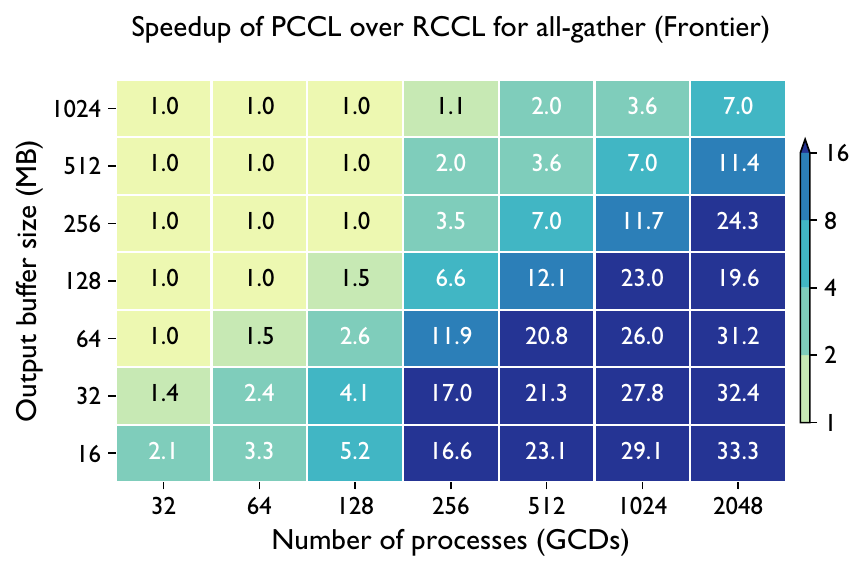}
  \includegraphics[width=0.325\textwidth]{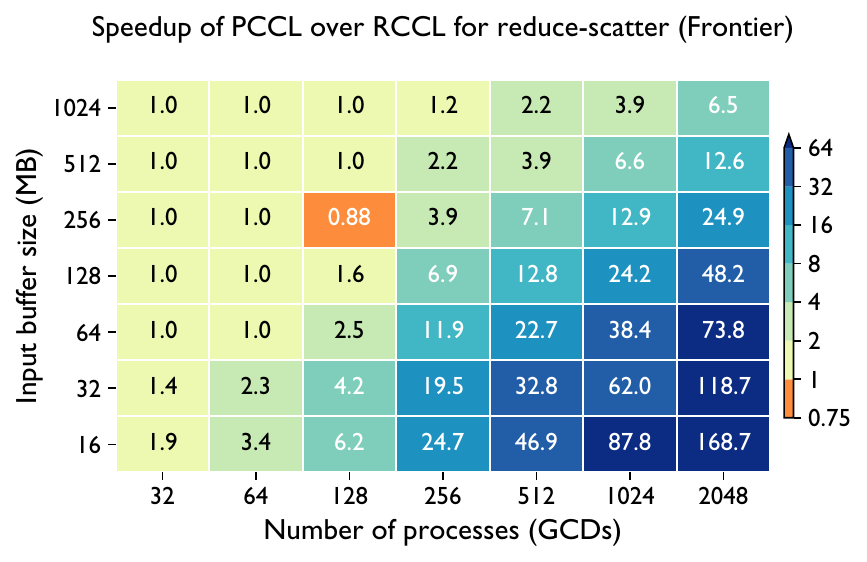}
  \includegraphics[width=0.325\textwidth]{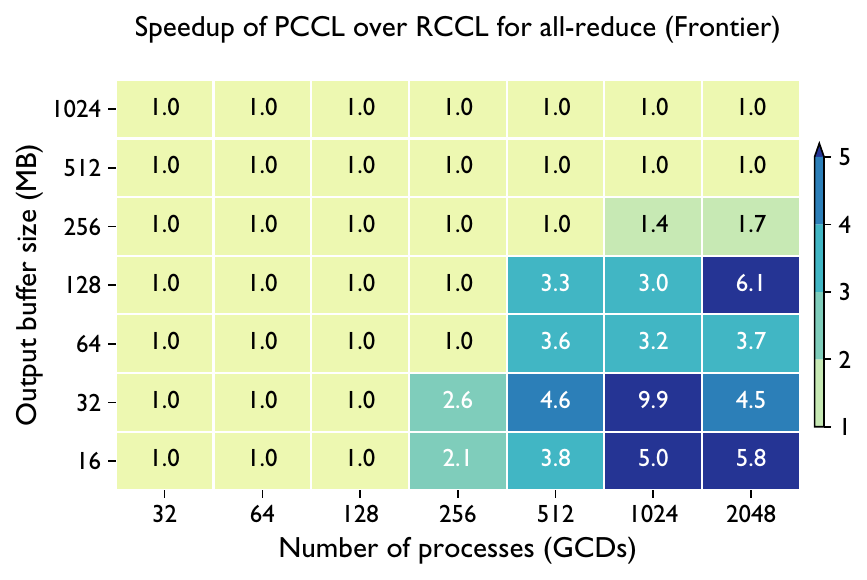}
  \caption{Heatmaps showing speedups from using PCCL with adaptive dispatching over RCCL for all-gather
(left), reduce-scatter (middle), and all-reduce (right) on Frontier. The speedup 
is shown as a function of per-process output/input buffer size (in MB) and process count.}
  \label{fig:heatmap-frontier}
\end{figure*}

We observe that RCCL and Cray-MPICH fell short of the ideal scaling line. Beyond 128
GCDs, RCCL's (green) execution time scales almost linearly with the number
of processes. Cray-MPICH (orange lines) demonstrates a similar pattern, with
performance dropping sharply as we scale to higher process counts. We attribute
the poor scaling of RCCL and Cray-MPICH to their reliance on the ring
algorithm, the latency of which grows linearly with the number of processes.

 
In contrast, PCCL (blue) maintains nearly flat scaling trends across
all message sizes, demonstrating significantly better scalability and
efficiency. We attribute PCCL's better performance to its hierarchical algorithms,
described in Section~\ref{sec:solutions}.
By using the ring algorithm within nodes (limited to eight
processes) and recursive doubling across nodes, PCCL bounds the latency
overhead that otherwise grows linearly in traditional ring-based
implementations. This design enables better scalability across large GPU
counts. The performance improvements of PCCL over RCCL and Cray-MPICH become
increasingly pronounced as we increase the number of processes. At 2048 GCDs,
PCCL all-gather achieves speedups ranging from $7$--$24\times$ over RCCL, and an even
larger $27$--$82\times$ over Cray-MPICH, depending on the message size. These
results highlight PCCL's ability to deliver highly efficient
communication at scale.

\begin{figure*}[t]
  \centering
  \includegraphics[width=0.49\textwidth]{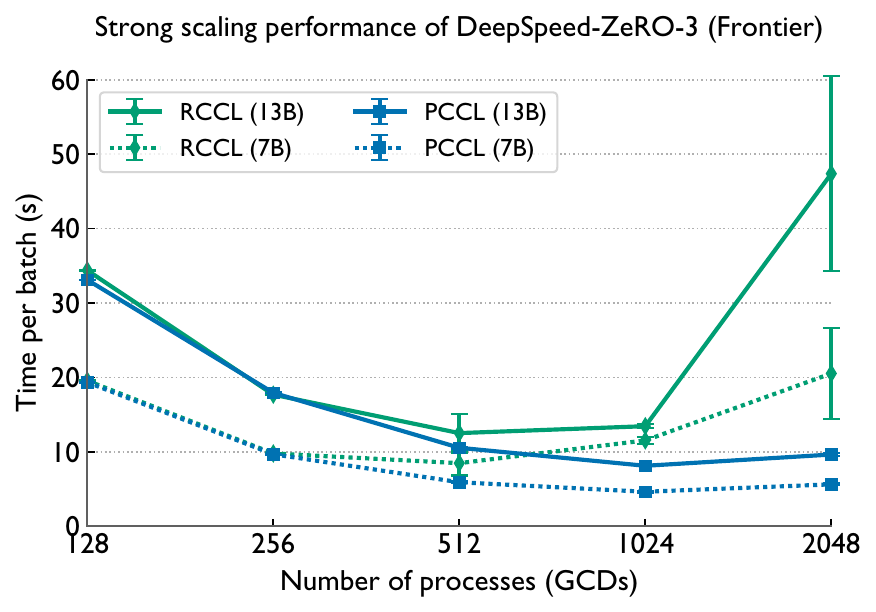}
  \includegraphics[width=0.49\textwidth]{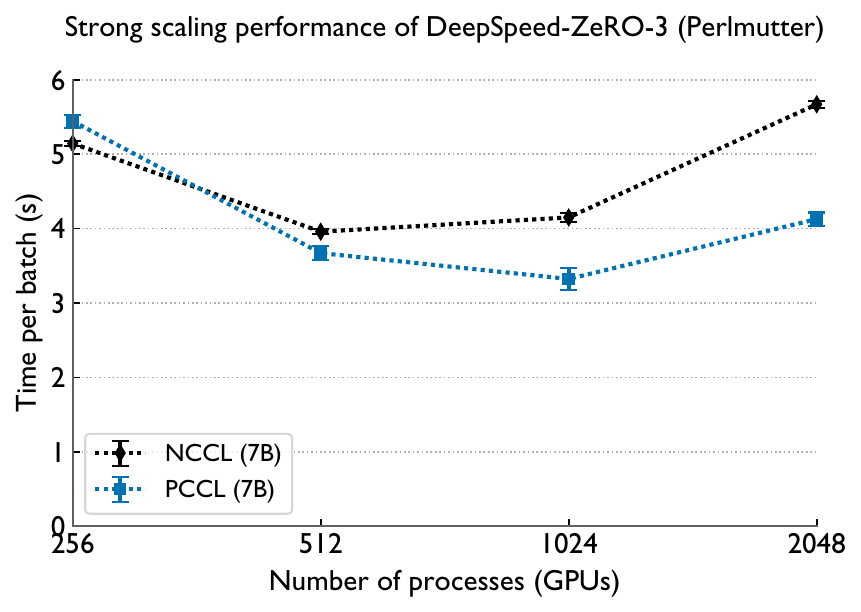}
  \caption{Strong scaling performance of DeepSpeed ZeRO-3 using RCCL or NCCL, and
PCCL, on Frontier (left) and Perlmutter (right) for two model sizes: GPT-3 7B
and GPT-3 13B.}
  \label{fig:e2e}
\end{figure*}

The performance trends for reduce-scatter on Frontier follow the same pattern
established by all-gather, as depicted in
Figure~\ref{fig:compare-libs-frontier} (middle). Both RCCL and Cray-MPICH fall
short of the ideal performance curve, and PCCL achieves significant speedups
over both libraries with increasing scale.
However, the speedups for all-reduce operations are comparatively smaller, as shown in Figure~\ref{fig:compare-libs-frontier} (right). Here, RCCL employs double-binary trees to achieve log-latency scaling, which explains its improved scalability over its own reduce-scatter or all-gather algorithms. 
Despite this, PCCL’s two-phase strategy demonstrates near-flat scaling, ultimately achieving significant speedups over RCCL at higher GCD counts.


Figure~\ref{fig:heatmap-frontier} shows the speedups of PCCL over RCCL for all-gather (left), reduce-scatter (middle), and all-reduce (right)
operations on Frontier, respectively, across a range of output buffer sizes and process counts. In the top-left regions of all three 
heatmaps, where we have large messages and small GPU counts, representing bandwidth-bound scenarios, PCCL's adaptive dispatching picks RCCL as the backend.
This is expected, as RCCL's flat ring algorithm can achieve higher bandwidth than PCCL's hierarchical two-phase strategy~\cite{choblueconnect2019}. 

In contrast, in the bottom-right corners of the heatmap, representing the
latency-sensitive regime with small messages and large GPU counts, PCCL
delivers substantial gains. 
On 2048 GCDs, for 16, 32, and 64 MB buffer sizes, PCCL achieves speedups of more than 30$\times$ and
50--150$\times$ over RCCL for all-gather and reduce-scatter, respectively.
To uncover the source of this performance improvement, we analyze the values of the Cassini NIC hardware 
counters for these runs.  Specifically, RCCL exhibits $200\times$ higher value for the 
\texttt{lpe\_net\_match\_overflow\_0} counter, compared to PCCL. 
According to the official Cassini documentation\footnote{\url{https://cpe.ext.hpe.com/docs/latest/getting_started/HPE-Cassini-Performance-Counters.html}}, this counter tracks the ``number of messages where 
payload data was delivered to a buffer on the overflow list because there was no match on the priority 
list'', noting that these messages ``incur higher cost because data must be copied from the overflow 
buffer''. While RCCL frequently triggers these expensive software-side 
copies, PCCL (by virtue of dissolving inter-node collective communication into MPI
point-to-point operations) maintains its communication on the hardware-accelerated ``priority list,'' ensuring 
zero-copy data movement.

For larger messages at 2048 GCDs, the speedups are comparatively smaller but
still significant-- 24.3 and 7.0$\times$ for all-gather, 24.9 and
6.5$\times$ for reduce-scatter on 256 and 1024 MB, respectively.  Similarly,
for all-reduce, PCCL achieves up to a 9.9$\times$ speedup over RCCL in regions
where latency dominates. These results underscore PCCL's strength in
latency-bound scenarios and highlight its ability to scale efficiently to
thousands of GPUs.

\subsection{Impact on DL Training Performance}

Finally, we examine how these communication gains translate into improvements in DL training performance at scale.
Figure~\ref{fig:e2e} (left) presents the batch times for strong scaling GPT-3-style transformer training on Frontier 
using the DeepSpeed ZeRO-3 framework~\cite{sc2020zero}. Green lines represent ZeRO-3 runs with RCCL, the default 
communication library and blue lines represent runs with all-gather and reduce-scatter collectives in ZeRO-3 issued with PCCL. At smaller 
scales (128 and 256 GCDs), both libraries perform comparably. However, as we scale further, PCCL begins to outperform RCCL. 
When scaling to 1024 GCDs, RCCL fails to maintain 
strong scaling and even exhibits increased batch times compared to 512 GCDs. In contrast, PCCL continues to scale efficiently, delivering 
a $2.5\times$ speedup for the 7B model and a $1.6\times$ speedup for the 13B model. Finally, at 2048 GCDs, although both libraries exhibit diminishing
returns in strong scaling efficiency, PCCL still achieves substantial speedups $3.3$--$4.9\times$ relative to RCCL.

We observed similar trends on Perlmutter, as shown in Figure~\ref{fig:e2e}
(right). For a 7B parameter model, at 256 GPUs, PCCL underperforms NCCL with
slowdown of $0.94\times$. However, as we scale to larger GPU counts, PCCL
begins to outperform NCCL--achieving a $1.07\times$ speedup at 512 GPUs and a
significantly higher $1.37\times$ speedup at 2048 GPUs.

\begin{figure}[h]
  \centering
  \includegraphics[width=\columnwidth]{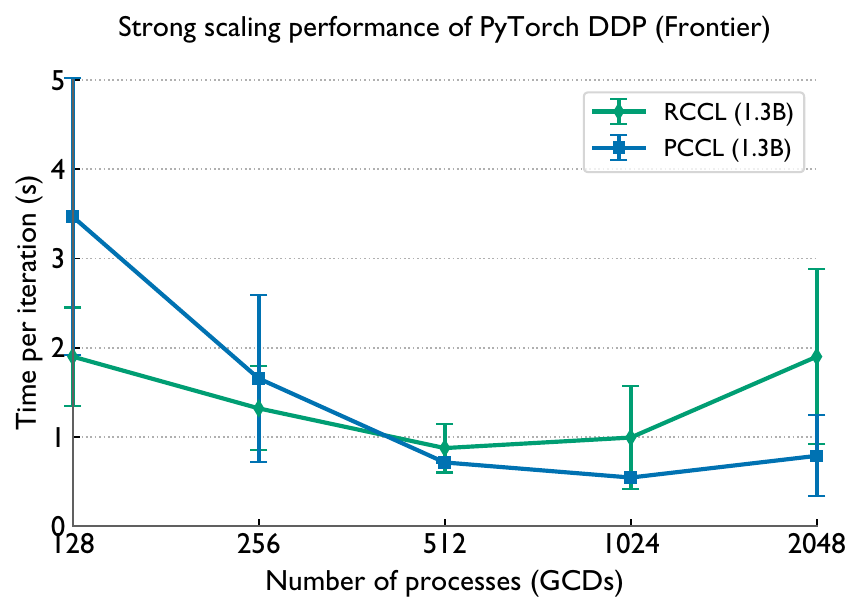}
\caption{Strong scaling performance of PyTorch DDP using RCCL and PCCL on Frontier 
for GPT-3 1.3B}
  \label{fig:e2e2}
\end{figure}

Figure~\ref{fig:e2e2} presents batch times for strong scaling GPT3-1.3B on Frontier using PyTorch DDP. 
At smaller scales, RCCL outperforms PCCL, with PCCL showing a $0.55\times$ and $0.80\times$ slowdown at 128 and 256 GCDs, respectively. However, at higher GCD counts, PCCL rapidly closes this performance gap and ultimately surpasses RCCL, achieving a $1.8\times$ and $2.4\times$ speedup over RCCL at 1024 and 2048 GCDs, respectively.
These results highlight PCCL's ability to deliver performance improvements for collective 
communication across multiple GPU architectures, and more importantly, translate those gains into significant
speedups for production DL applications. 

\section{Related Work}
Optimizing collective communication has been a long-standing challenge in high
performance computing (HPC) and parallel computing and has been the focus on
extensive research. Thakur et al.'s seminal work focuses on optimizing a
plethora of collective operations including all-gathers and reduce-scatters in
MPICH~\cite{thakurimproving2003}. The authors explore the design space of
several algorithms for each collective, and provide guidelines for selecting
the most appropriate algorithm for different scenarios. In contrast, our work
focuses on large-scale deep learning workloads and with messages sizes in the
tens to hundreds of megabytes.

Patarsuk et al.~propose the bandwidth-optimized ring algorithm for all-reduce
operations~\cite{ring-all-reduce}. Graham et al.~develop optimize MPI
collective to effectively exploit shared memory on multi-core systems. Chan et
al.~\cite{chan2006simultaneouscollectives} develop highly optimized collectives
for the IBM Blue Gene/L, exploiting unique properties of the
system~\cite{grahammulticorempi2008}.  Kandalla et al.~develop a scalable
multi-leader hierarchical algorithm for
all-gather~\cite{kandalla2009multileader}.  Note that the focus of these works
is on optimizing the performance of collective communication in traditional HPC
workloads with small message sizes, and not on the large message sizes inherent
to deep learning. 

De Sensi et al.~study the performance of NCCL, RCCL and Cray-MPICH across
various state-of-the-art supercomputers across a variety of latency and
bandwidth bound scenarios~\cite{sensiexploring2024}.  Cho et al.~propose a
multi-level hierarchical ring algorithm for all-reduce and study the tradeoff
of bandwidth and latency between the flat and hierarchical ring
algorithms~\cite{choblueconnect2019}. In this work, we build on this and
exploit more latency optimal algorithms such as recursive doubling/halving in the
inter-node levels of the hierarchy and also demonstrate how this design can be
utilized to load-balance network traffic across NICs. Note that similar
hierarchical designs have been explored in other works as
well~\cite{hashmidesigning2018, nguyen2hierarchical2018}.

Hidayetoglu et al.~present HiCCL, a hierarchical collective communication
library targeting the same systems as our work -- Perlmutter and
Frontier~\cite{hidayetoglu2024hicclhierarchicalcollectivecommunication}. While HiCCL demonstrates performance advantages over
NCCL and RCCL, these gains are primarily observed at smaller node counts. As
shown in their evaluation, HiCCL's performance matches or falls below vendor
libraries at larger node counts. In contrast, PCCL is specifically designed and
optimized for the large-scale regime, consistently outperforming vendor
libraries at thousands of GPUs. Thus, HiCCL and PCCL address complementary
performance regimes.

Cai et al.~develop a systematic theoretical approach to synthesize novel
communication algorithms for optimizing collective communication on a
particular topology~\cite{cai2021synthesizing}.  Cho et al.~develop a strategy
to maximize the overlap of a tree-based all-reduce with the computation in
neural network training~\cite{cho2023logical}. There is also a body of work
focused on exploiting data compression to minimize communication overheads in
distributed deep learning. For example, Feng et al.~optimize all-to-all
communication in recommendation model training via a novel error-bounded
compression algorithm~\cite{feng2024accelerating}. Huang et al.~develop hZCCL,
a communication library that enables collective operations on compressed
data~\cite{huang2024hzccl}. Zhou et al.~develop a GPU-based compression scheme
for all-gathers and reduce-scatters~\cite{zhou2023accelerating} and optimize
FSDP~\cite{fsdp} training at scale.

In the context of algorithm selection and optimization, Liu et al.~develop
parameterized ring and recursive algorithms for GPU-to-GPU collectives, and
introduce a simulation technique for parameter auto-tuning~\cite{liu2025parameterized}.
Wilkins et al.~develop ACCLAiM, which uses machine learning for auto-tuning MPI
collective communication~\cite{Wilkins2022ACCLAiMAT}, an approach conceptually
related to our SVM-based adaptive dispatching mechanism. Hunold et al.~present an auto-tuning framework that uses regression models to predict the fastest
algorithm for different collective routines~\cite{hunold:cluster2020}.
Venkata et al.~develop UCC, a unified collective communication library with
pluggable transport layers supporting CPUs, GPUs, and DPUs, targeting broad
applicability across programming models and
hardware~\cite{DBLP:conf/hoti/VenkataPLBALBDS24}.


\section{Conclusion}
\label{sec:conc}
We investigated the current state of performance of collective communication
routines in several state-of-the-art communication libraries.  Specifically, we
focused on the all-gather, reduce-scatter, and all-reduce collectives, which
are commonly used in distributed deep learning workloads -- both during
training and inference.  We evaluated the performance of Cray-MPICH, RCCL, and
NCCL on leadership-class GPU supercomputers, and identified critical
performance bottlenecks that limit their scaling to large GPU counts.

In order to address these limitations, we introduced PCCL, a scalable,
performant and portable communication library with highly optimized
implementations of all-gather, reduce-scatter, and all-reduce operations.  PCCL
combines three key ideas:~(1)~hierarchical two-level collective designs that
better exploit the available hardware, (2) latency-optimal recursive
doubling/halving implementations for inter-node communication, and (3) an
SVM-based adaptive dispatcher that dynamically selects the most performant
backend based on message size and GPU count.  We demonstrated significant
performance improvements over all three libraries, both in collective
communication benchmarks as well as production DL training. In future work, we
plan to benchmark PCCL on clusters with InfiniBand interconnects to evaluate
its performance beyond Slingshot-based systems.  As model sizes and system
scales continue to grow, we believe that intelligent, architecture-aware
collective communication will be a central enabler of efficient large-scale AI
workloads.

\section*{Acknowledgment}
This material is based upon work supported in part by the National Science
Foundation under Grant No.~2047120. We thank Joy Kitson of the Parallel
Software and Systems Group for providing feedback on the paper and for
improving the heatmap figures in the paper.

This research used resources of the Oak Ridge Leadership Computing Facility at
the Oak Ridge National Laboratory, which is supported by the Office of Science
of the U.S.~Department of Energy (DOE) under Contract No.~DE-AC05-00OR22725.
This research also used resources of the National Energy Research Scientific
Computing Center (NERSC), a U.S.~DOE Office of Science User Facility, operated
under Contract No.~DE-AC02-05CH11231 using NERSC award DDR-ERCAP0034262.

\bibliographystyle{IEEEtran}
\bibliography{./bib/cite,./bib/pssg}

\clearpage
\appendices
\def\IPDPSADINCLUDED{1}
\appendixAD

\section{Overview of Contributions and Artifacts}

\subsection{Paper's Main Contributions}

\begin{description}
    \item[$C_1$] Analyzing the limitations of existing communication libraries, Cray-MPICH and RCCL, for all-gather, reduce-scatter, and all-reduce
    collectives in parallel deep learning workloads.
    \item[$C_2$] Developing optimized implementations of these collectives in PCCL (the communication library proposed in this work),
    with a focus on effectively utilizing system resources and ensuring scalability in latency bound scenarios. PCCL also includes a
   machine learning algorithm that selects the best communication backend at runtime based on GPU count and message size.
    \item[$C_3$] Conducting benchmarking of large-scale LLM training workloads using both DeepSpeed ZeRO-3 and PyTorch DDP
    to validate the practical benefits of our optimizations, demonstrating significant speedups in training throughput.
\end{description}

\subsection{Computational Artifacts}

$A_1$ (\href{https://github.com/hpcgroup/pccl-reproducer-ipdps26}{PCCL}) is our implementation of the accelerated all-gather, reduce-scatter, and all-reduce collectives proposed in this work.

$A_2$ (\href{https://github.com/axonn-ai/nanoGPT/tree/new-collectives}{nanoGPT, DeepSpeed ZeRO-3 branch}) is a codebase for training large language models (LLMs) using the DeepSpeed parallel training framework. It supports both NCCL/RCCL and PCCL as communication backends.

$A_3$ (\href{https://github.com/axonn-ai/nanoGPT/tree/pccl-ddp}{nanoGPT, PyTorch DDP branch}) is a codebase for training LLMs using PyTorch DDP, used for benchmarking of PCCL's all-reduce against RCCL on Frontier.

\section{Artifact Identification}

\newartifact

\artrel

Artifact $A_{1}$ is PCCL, the communication library proposed in this work, featuring optimized implementations of all-gather,
reduce-scatter, and all-reduce operations. This artifact includes scripts for benchmarking these collectives in isolation and for comparing their performance
against NCCL, RCCL, and Cray-MPICH. It also provides scripts for building PCCL and its dependencies, as well as for running the comparison
benchmarks on Perlmutter and Frontier.

\artexp

Reproducing $A_{1}$ allows us to reproduce all plots in the paper except Figures~12 and~13.

\arttime

\textbf{Artifact Setup:} Building the artifact takes less than five minutes on both Perlmutter and Frontier.

\textbf{Artifact Execution:} For a given process count, a communication library
(RCCL/NCCL, Cray-MPICH, or PCCL), and a collective (all-gather/reduce-scatter/all-reduce), profiling times across all message sizes (1\,MB to 1024\,MB) takes around 2 minutes. To complete the full sweep, one must launch jobs for all three libraries, the three collectives, as well as all
node counts (4--256 nodes on Frontier, and 8--512 on Perlmutter), and multiple trials (atleast 10) to account for performance variability. While these jobs can execute in parallel, the exact execution time depends on the job scheduler. In practice, 24 hours are sufficient to gather all data points.

\textbf{Artifact Analysis:} Analyzing the outputs of this artifact involves running grep commands on the outputs, which overall take
less than 10 seconds.

\artin

\artinpart{Hardware}

This artifact has been tested for correctness and performance on Perlmutter, which has NVIDIA A100 GPUs, and Frontier, which
has MI250X GPUs. We expect it to run on any cluster that supports PyTorch (and its distributed communication backend) and a
performant distribution of MPI (like Cray-MPICH on Cray machines).

\artinpart{Software}

\begin{center}
    \begin{tabular}{rll}
    \toprule
    Software Name   &  Version & Comments \\
    and URL         &          &  \\
    \midrule
    \href{https://github.com/hpcgroup/pccl-reproducer-ipdps26}{PCCL} & 0.0.1 & N/A  \\
    \midrule
    \href{https://cpe.ext.hpe.com/docs/24.03/mpt/mpich/index.html}{Cray-MPICH} &  8.1.31 (Frontier)   & Available as  \\
                                                                               &  8.1.30 (Perlmutter) & modules   \\
    \midrule
    \href{https://pytorch.org/get-started/locally/}{PyTorch} &  2.6.0 & Ships with  \\
                                                             &        & RCCL/NCCL \\
    \midrule
    \href{https://github.com/ROCm/aws-ofi-rccl}{AWS Plugin} & 1.4 & Only for Frontier \\
    \midrule
    \href{https://www.amd.com/en/products/software/rocm.html}{ROCm} & 6.4.1 & Available as a \\
    & & module on Frontier \\
    \midrule
    \href{https://developer.nvidia.com/cuda-toolkit}{CUDA} & 12.4 & Available as a \\
    & & module on Perlmutter \\
    \bottomrule
    \end{tabular}
\end{center}

\artinpart{Datasets / Inputs}

Input data for the communication benchmarks is generated on the fly (randomly initialized GPU tensors) in the scripts.

\artinpart{Installation and Deployment}

Bash scripts in the \texttt{scripts/} folder install this artifact and its dependencies on both machines.

\artcomp

This is the task workflow, assuming the artifact has been built successfully. One can complete all of these tasks using
variations of the sample bash script \texttt{run\_raw\_collectives\_benchmark.sh} provided for each machine in
the \texttt{scripts} folder. The scripts can be modified to change the collective to be benchmarked (between all-gather,
reduce-scatter, and all-reduce) and the communication library (NCCL/RCCL, Cray-MPICH, PCCL).

\begin{description}
    \item[$T_1$] Benchmark all-gather times for NCCL/RCCL on all node counts (4--256 nodes on Frontier and 8--512 nodes on Perlmutter).
    Repeat ten times.
    \item[$T_2$] Benchmark all-gather times for Cray-MPICH on all node counts. Repeat ten times.
    \item[$T_3$] Benchmark all-gather times for PCCL on all node counts. Next, change the algorithm to recursive and run the benchmarks again. Repeat ten times.
    \item[$T_4$] Repeat $T_1$--$T_3$ for reduce-scatter.
    \item[$T_5$] Repeat $T_1$--$T_3$ for all-reduce.
\end{description}

\artout

The outputs of these tasks are execution times for different message sizes and node counts.

\newartifact

\artrel

Artifact $A_2$ is a codebase for training an LLM on multiple GPUs using the DeepSpeed parallel training framework.
It supports PCCL, NCCL, and RCCL as communication libraries for all-gathers and reduce-scatters. We use
this artifact to generate the results shown in Figure~12.

\textbf{Note:} One must use the \texttt{pccl-zero-3} branch of the
\href{https://github.com/axonn-ai/nanoGPT/tree/pccl-zero-3}{GitHub repo} for this artifact.

\artexp

In Figure~12 we compare the batch iteration times for training 7B and 13B parameter LLMs on Frontier and Perlmutter
respectively. For each GPU count we first run training with RCCL/NCCL (the default communication library in DeepSpeed)
and then repeat with PCCL. Given that PCCL's speedups over RCCL/NCCL increase with process count, we expect the same
trend in end-to-end training, i.e., PCCL should provide increasing speedups with more GPUs.

\arttime

Each run (a given model size, process count, and communication library) requires approximately 10 minutes.

\artin

\artinpart{Hardware}

Same as $A_{1}$.

\artinpart{Software}

All dependencies of $A_1$ are required. Additionally, the following Python packages are needed:
\texttt{transformers}, \texttt{datasets}, \texttt{tiktoken}, and \texttt{tqdm}.
All can be installed with \texttt{pip install <package-name>}.

\artinpart{Datasets / Inputs}

This artifact uses the OpenWebText dataset. Scripts to download and prepare the dataset are provided in
\texttt{data/openwebtext/}. Note that we use a small subset of the dataset for benchmarking (~5\%), so the preparation steps are 
fast (less than 30 minutes). The scripts also support using the full dataset if desired.

\artinpart{Installation and Deployment}

Bash scripts in the \texttt{scripts/} folder install this artifact and its dependencies on both machines.

\artcomp

This is the task workflow, assuming the artifact has been built successfully. One can complete all tasks using
variations of the sample bash script \texttt{run\_deepspeed\_benchmark.sh} provided for each machine in
the \texttt{scripts} folder. The scripts can be modified to change the model size (7B or 13B) and the
communication library (NCCL/RCCL or PCCL).

\begin{description}
    \item[$T_1$] Benchmark the 7B model on all node counts (16--256 nodes on Frontier and 64--512 on Perlmutter) using RCCL/NCCL. Repeat 5 times.
    \item[$T_2$] Benchmark the 7B model on all node counts using PCCL. Repeat 3 times.
    \item[$T_3$] Repeat $T_1$--$T_2$ for the 13B model.
\end{description}

Note: the 13B model runs out of memory at almost all GPU counts on Perlmutter.

\artout

The outputs are batch iteration times for each run. These are used as data points for Figure~12.

\newartifact

\artrel

Same nanoGPT codebase as $A_{2}$ (\href{https://github.com/axonn-ai/nanoGPT/tree/pccl-ddp}{\texttt{pccl-ddp} branch}),
but trains with PyTorch DDP and all-reduce to produce Figure~13.

\artexp

Figure~13 shows batch iteration times for training a GPT-3 1.3B model on Frontier using PyTorch DDP with RCCL vs.\ PCCL.
Because DDP relies on all-reduce (rather than all-gather/reduce-scatter), this experiment validates PCCL's all-reduce
performance at scale. The crossover point where PCCL surpasses RCCL is expected to occur at large GPU counts ($\geq$1024 GCDs).

\arttime

$\sim$10~min per run (process count $\times$ communication library).

\artin

\artinpart{Hardware}
Same as $A_{1}$.

\artinpart{Software}
Same as $A_{2}$.

\artinpart{Datasets / Inputs}
Same as $A_{2}$.

\artinpart{Installation and Deployment}
Same as $A_{2}$; use the \texttt{pccl-ddp} branch.

\artcomp

Use \texttt{run\_frontier.sh} (in the \texttt{scripts/} folder) and modify the communication library flag.

\begin{description}
    \item[$T_1$] Benchmark GPT-3 1.3B on Frontier (128--2048 GCDs) using RCCL. Repeat 5 times.
    \item[$T_2$] Repeat $T_1$ with PCCL. Repeat 5 times.
\end{description}

\artout

The outputs are batch iteration times per run, used to produce Figure~13.

\end{document}